\documentclass[aps,prl,twocolumn,superscriptaddress,showpacs,floatfix]{revtex4-1}

\usepackage[english]{babel}
\usepackage{graphicx}
\usepackage{multirow}
\usepackage{amsmath,amssymb,amsfonts}
\usepackage{amsthm}
\usepackage{mathrsfs}
\usepackage{xcolor}
\usepackage{textcomp}
\usepackage{upgreek}
\usepackage{booktabs}
\usepackage{bm}%
\usepackage{hyperref}
\usepackage[protrusion=true,expansion=true]{microtype}%
\usepackage{etoolbox}
\AtBeginEnvironment{thebibliography}{\raggedright}%

\newcommand{\equ}[1]{Eq.~(\ref{#1})}

\begin{document}

\title{Spatially Resolving the Pre-Thermal Anatomy of a Driven Bosonic Fluid}

\author{Shantam Ravan}
\affiliation{Department of Physics, Harvard University, 12 Oxford Street, Cambridge, MA 02138, USA}
\affiliation{Department of Physics, University of Maryland, College Park, 4296 Stadium Dr, College Park, MD 20742, USA}

\author{Aaron M\"uller}
\affiliation{Department of Physics, ETH Zurich, Otto-Stern-Weg 1, 8049 Z\"urich, Switzerland}

\author{Johannes Cremer}
\affiliation{Department of Physics, Harvard University, 12 Oxford Street, Cambridge, MA 02138, USA}
\affiliation{Quantum Technology Center, University of Maryland, College Park, 8228 Paint Branch Dr, College Park, MD 20742, USA}

\author{Jonathan Curtis}
\affiliation{Department of Physics, ETH Zurich, Otto-Stern-Weg 1, 8049 Z\"urich, Switzerland}

\author{Daniel Fernandez}
\affiliation{Department of Physics, Harvard University, 12 Oxford Street, Cambridge, MA 02138, USA}

\author{Ronald Walsworth}
\affiliation{Quantum Technology Center, University of Maryland, College Park, 8228 Paint Branch Dr, College Park, MD 20742, USA}
\affiliation{Department of Physics, University of Maryland, College Park, 4296 Stadium Dr, College Park, MD 20742, USA}
\affiliation{Department of Electrical and Computer Engineering, University of Maryland, College Park, 8228 Paint Branch Dr, College Park, MD 20742, USA}

\author{Eugene Demler}
\affiliation{Department of Physics, ETH Zurich, Otto-Stern-Weg 1, 8049 Z\"urich, Switzerland}

\author{Amir Yacoby}
\affiliation{Department of Physics, Harvard University, 12 Oxford Street, Cambridge, MA 02138, USA}

\begin{abstract}
Understanding how coherently driven quantum many-body systems redistribute energy prior to thermal equilibrium remains a central challenge in many-body physics. Here, we utilize nitrogen-vacancy (NV) magnetometry to perform micron-scale spatial imaging of room-temperature magnon dynamics in a yttrium iron garnet (YIG) thin film. We resolve a hierarchy of discrete parametric scattering events that serve as deterministic stepping stones toward thermalization. By applying a two-tone wave-mixing protocol, we first isolate the elementary four-magnon interaction and extract its coupling strength via the spatial growth of the scattering product. We then drive the system with an intense single-frequency excitation near ferromagnetic resonance, revealing that magnon-magnon interactions trigger a spontaneous, multi-generation scattering cascade. We demonstrate that in each generation, the dominant scattering channels correspond to one of the out-scattered magnons being in the slow magnon regime, reminiscent of the enhancement of optical nonlinearities in slow light systems. We capture this dynamics quantitatively using a near field magnonics framework and extract the cascade order and nonlinear coefficients directly from power-dependent frequency shifts. By revealing the multi-stage dynamical process through which monochromatic injected magnons evolve toward equilibrium, our work establishes spatially resolved magnonics as a powerful platform for visualizing non-equilibrium many-body kinetics.
\end{abstract}

\maketitle

\section{Introduction}\label{introduction}

Relaxation of interacting many-body systems to equilibrium is one of the longest-standing fundamental problems in physics. Since Boltzmann introduced the idea of ergodicity, the general picture has been that systems relax to local equilibrium on microscopic timescales, retaining only coarse-grained memories of density and temperature, before hydrodynamic variables relax toward global equilibrium~\cite{polkovnikov_colloquium_2011}. Over the years, several exceptions to this canonical scenario have been identified, including integrable systems~\cite{rigol_thermalization_2008} and many-body localization~\cite{nandkishore_many-body_2015}. In these cases, strong disorder or kinematic constraints strictly prevent the system from exploring phase space, effectively halting thermalization.

In this paper, we report the experimental observation of a distinct relaxation scenario in which dynamics proceed through a cascade of scattering processes at well-defined momenta. Many earlier works studying the redistribution of energy towards equilibrium have commonly used Wave Turbulence Theory (WTT) to model a smooth, stochastic process~\cite{zakharov1992kolmogorov, nazarenko2011wave}. This includes work on quark-gluon plasmas~\cite{berges_thermalization_2021} and sea-state spectra~\cite{hasselmann_non-linear_1962, annenkov_spectral_2018}. However, concrete examples from ultracold atomic gases and photonics have hinted that coherence flows through discrete, phase-matched channels long before it thermalizes. In Bose--Einstein condensates (BECs), periodic modulation triggers parametric instabilities that emit `firework' patterns of excitations at well-defined momenta~\cite{clark_collective_2017,fu_density_2018}, generate Faraday waves and granulation~\cite{nguyen_parametric_2019}, and form density patterns stabilized by non-equilibrium fixed points~\cite{fujii2024stable, oberthaler2025pattern}. Similarly, in semiconductor microcavities, driven exciton-polaritons scatter into specific signal and idler modes governed by strict phase-matching rules~\cite{savvidis_angle-resonant_2000,baumberg_parametric_2005}.

We utilize magnons, the bosonic quanta of spin waves, in yttrium iron garnet (YIG) to demonstrate this novel relaxation phenomenon in a room-temperature solid-state system. YIG supports strong nonlinear interactions with exceptionally low damping~\cite{serga_yig_2010,chumak_magnon_2015}, enabling landmark observations of magnon Bose--Einstein condensation~\cite{demokritov_boseeinstein_2006,serga_boseeinstein_2014} and thermalization~\cite{demidov_thermalization_2007,du_control_2017}, and allowing us insight into magnon-magnon scattering dynamics. While these have been previously characterized by Brillouin light scattering~\cite{sebastian_micro-focused_2015,schultheiss_direct_2012}, nitrogen-vacancy (NV) diamond magnetometry~\cite{van_der_sar_nanometre-scale_2015,bertelli_magnetic_2020} has proven to be a powerful tool for studying magnonics, especially given its ability to measure the absolute magnetic field, with both high spatial resolution and wide field-of-view across a sample of interest. 

In our work, we leverage NV magnetometry to directly image the scattering dynamics of magnons in a YIG thin film. We first isolate a single scattering process via stimulated wave mixing, extracting the four-magnon interaction strength from its spatial growth profile. We then track the spontaneous cascade of magnon scattering processes under strong driving, discovering that the system populates a hierarchy of discrete, power-tunable resonances. These are cascaded parametric modes where a coherent drive spawns successive generations of daughter magnonic waves. Standard perturbative approaches like S-theory (the mean-field theory of parametrically excited waves)~\cite{lvov_wave_1994, rezende1990spin} predict the primary instability threshold but fail to capture the multi-stage kinetics preceding thermalization. We develop a new model which demonstrates that the buildup of these modes is governed by dynamic selection rules that favor $\bm k$-values where the group velocity vanishes, effectively isolating the instability. At these slow magnon points the magnon energy density is compressed and the four-magnon interaction amplified, in analogy to the enhancement of optical nonlinearities in slow-light media~\cite{soljacic_photonic_2002, hamachi_slow_2009, boyd_material_2011}. By correlating spatial growth with mean-field frequency shifts and modeling this process using conjugate wave optics and Bogoliubov theory, we identify these resonances as the nonlinear pathways (the stepping stones) by which the coherent fluid restructures itself before equilibration. Our results establish thin-film magnonics and NV imaging as a platform for visualizing the deterministic anatomy of thermalization in driven quantum many-body systems.

\section{Results}\label{results}

\subsection{Single Frequency Magnon Generation and Detection}\label{results:stimulated_four_magnon_scattering}

\begin{figure}[htbp]
\centering
\includegraphics[width=\columnwidth]{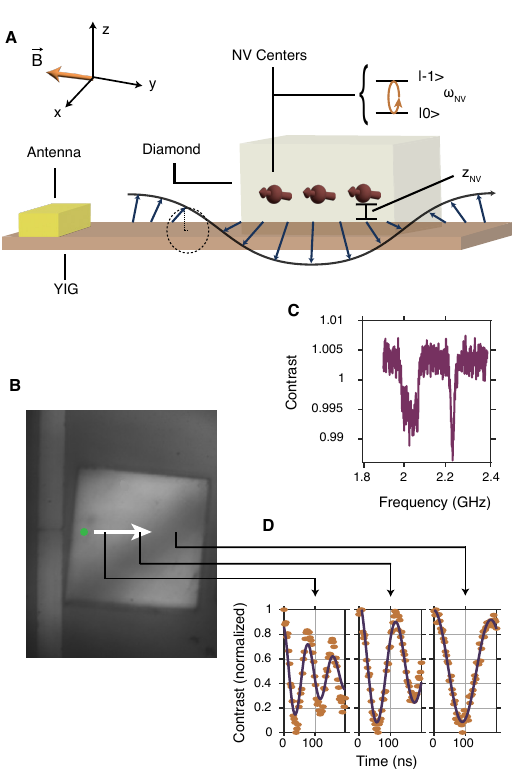}
\caption{Experimental setup and single-frequency magnon detection. \textbf{(A)} Schematic diagram of the experimental geometry showing the YIG thin film, diamond chip with NV centers, microwave stripline antenna, and magnetic bias field orientation. Spin waves in the YIG film are a result of the precession of individual magnetic dipoles around the equilibrium magnetization, as indicated by the dashed circle.  \textbf{(B)} Optical camera image of the diamond chip placed on the surface of the YIG adjacent to the stripline antenna. The green dot indicates the laser spot of the confocal microscope, used for NV excitation, which is scanned across the diamond chip to probe magnons at different positions along the YIG film. The white arrow indicates the scan direction. \textbf{(C)} Example NV optically detected magnetic resonance (ODMR) spectrum acquired with a 20\,mT bias field. There is a broad dip in NV photoluminescence (PL) around 2.05\,GHz due to nonlinear magnon generation, paired with a sharp PL dip at the NV resonance frequency. \textbf{(D)} Coherent NV Rabi oscillation data taken at a series of positions on the diamond chip and hence the YIG film, as indicated by the black arrows. The stripline is driven with a single microwave frequency tone at the NV resonance and the resulting NV PL signal is fit to an oscillating sinusoid. As the chip does not sit fully flat against the YIG film, the NV/YIG separation slightly increases as one moves further from the stripline, leading to a weaker magnon-driven stray field and lower Rabi frequency (see Supplementary Information for technical details).}
\label{fig:paper_fig_1}
\end{figure}

To directly image the spatial evolution of the magnon gas, we employ a dense ensemble of nitrogen-vacancy (NV) centers in close proximity to a room-temperature, 100\,nm-thick YIG thin film (Fig.~\ref{fig:paper_fig_1}(A,B)). The diamond chip (50\,$\upmu$m $\times$ 50\,$\upmu$m), that hosts the NV ensemble, is positioned with an average distance of $z_{\rm NV} \approx 300$~nm above the YIG surface, and the NVs are probed with a custom-built setup including a confocal microscope and microwave antenna (see Methods), ensuring non-invasive sensing of the local dipolar fields from the YIG film while maintaining sub-micron spatial resolution~\cite{ku_imaging_2020, barry_sensitivity_2020}. An external magnetic bias field $H_{\rm ext}$ is applied at $\theta_{\rm NV} = -71^{\circ}$ relative to the surface normal, to enable probing of one orientation class of NVs within the diamond chip and to orient the magnetization for the excitation of Damon-Eshbach YIG spin waves~\cite{kalinikos_theory_1986}.

We utilize two complementary NV sensing modalities to map the spectral and spatial magnon characteristics of the excitations. First, we employ NV optically detected magnetic resonance (ODMR) spectroscopy to identify the local magnon spectrum. As shown in Fig.~\ref{fig:paper_fig_1}(C), the ODMR spectrum with a 20~mT bias field reveals a sharp resonance at 2.2~GHz, corresponding to direct excitation of magnons at the NV resonance frequency. Crucially, we also observe a broad, anomalous magnon absorption feature centered at 2.05~GHz. As we discuss in the following sections, this signature arises not from direct magnon driving, but from the nonlinear generation of cascaded magnon scattering products.

Second, to quantify the local spin wave amplitude, we drive coherent Rabi oscillations of the NV sensor at the magnon frequency (Fig.~\ref{fig:paper_fig_1}(D)). By scanning the confocal volume of probed NVs away from the microwave antenna, we observe a spatial decay in the Rabi frequency $\Omega_R$, which is proportional to the local microwave field amplitude $B_{\rm mw} \propto \Omega_R$. This calibration confirms our ability to spatially resolve the propagation of coherent magnon excitations over finite distances, a prerequisite for imaging the kinetic buildup of the cascade of magnon scattering processes~\cite{zhou_magnon_2021}.

\subsection{Spatially Resolving the Nonlinear Four-Magnon Interaction Strength}
\label{results:nonlinear_interaction_strength}

We begin by isolating the elementary four-magnon scattering process using a stimulated wave-mixing protocol. These magnon-magnon processes, which underlie the second-order Suhl instability~\cite{suhl_nonlinear_1957}, have recently emerged as a powerful tool for studying magnon dynamics, including creating spin-wave frequency combs~\cite{hula_spin-wave_2022} and coherently generating target magnon modes~\cite{carmiggelt_broadband_2022}. However, these efforts largely exploit the nonlinearity as a tool, rather than explore the intricate many-body dynamics that it produces. While Brillouin light scattering experiments have directly observed these scattering processes at finite wavevectors~\cite{wettling_light_1983, schultheiss_direct_2012}, quantitative measurements of the microscopic interaction vertex $g$ remain rare~\cite{lvov_wave_1994}.

\begin{figure*}[t]
    \centering
    \includegraphics[width=\textwidth]{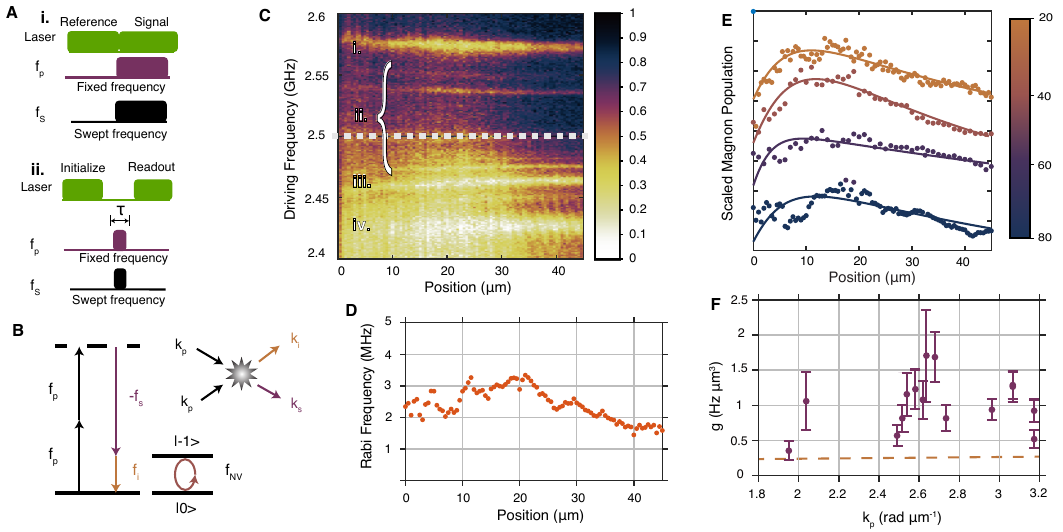}
    \caption{\textbf{(A)} (i) Pulse sequence to acquire data in (C): a standard NV ODMR sequence with a microwave signal at fixed frequency $f_p$ and an additional driving tone at swept frequency $f_s$. (ii) Pulse sequence to acquire data in (D). The length of the microwave signal for both $f_p$ and $f_s$ is swept to drive Rabi oscillations between the NV spin states. \textbf{(B)} Schematic of the four magnon NV mixing process. The left side indicates that the consumption of two $f_p$ magnons and the emission of $f_s$ and $f_i$ magnons conserves energy. If the $f_i$ idler magnon is at the NV spin resonance, then the NV ensemble will be coherently driven. \textbf{(C)} Example measured NV ODMR spectrum along the diamond chip and hence the YIG film, as well as swept drive frequency $f_s$, under the application of the microwave dual-tone drive at fixed bias magnetic field (16\,mT). The zero to one scale of this plot indicates normalized ODMR photoluminescence (PL) contrast (see Supplementary Information). Four ODMR resonances are visible in the spectrum, labeled as (i), (ii), (iii) and (iv). (i) and (iii) are NV PL dips due to the $\pm$ first order stimulated four-magnon scattering processes ($2f_p - f_s$ or $2f_s - f_p$ respectively) being resonant with the NV transition. (ii) represents a set of higher order stimulated four-magnon scattering processes ($(n+1)f_p - nf_s$) on spin resonance with the NV transition. (iv) is direct microwave field driving of the NV spin resonance ($f_{\rm NV}$). White dashed line indicates the value of $f_p$. \textbf{(D)} NV Rabi frequency data taken as a function of position during a lateral scan of the diamond chip and hence the YIG film, along the direction of the arrow in Fig.~\ref{fig:paper_fig_1}(B). The YIG film is driven with a two-tone microwave signal, such that the idler frequency $f_i$ is resonant with the NV ODMR transition, thereby driving Rabi oscillations (similar to those seen in Fig.~\ref{fig:paper_fig_1}(D)). These Rabi oscillations are fit to a decaying sinusoid whose frequency is extracted and plotted here. \textbf{(E)} Idler magnon mode population density $\langle n_{\bm k} \rangle$ as a function of position, for $f_p$ values detuned 20--80\,MHz from $f_{\rm NV}$ at a bias field of 16\,mT. $\langle n_{\bm k} \rangle$ values are calculated from experimentally-determined Rabi frequencies. From top to bottom, the datasets correspond to detunings of 20, 40, 60, and 80\,MHz. Solid lines are fits of the data to the coupled-mode theory (\equ{eq:idler_envelope}). \textbf{(F)} Extracted four-magnon couplings $g^{pp}_{si}$ (purple dots) from the model in \equ{eq:idler_envelope} as a function of magnon wavenumber $k_p$ for a series of bias fields from 16\,mT to 20\,mT and $f_p$ values detuned similarly as in (E), compared with the T-matrix prediction (orange dashed line).}
    \label{fig:paper_fig_2}
\end{figure*}

Here, we implement a spatially resolved scheme, using NV magnetic imaging (Fig.~\ref{fig:paper_fig_2}(A)), similar to that employed in Ref.~\cite{carmiggelt_broadband_2022}. This approach coherently generates and images a specific scattering product, allowing us to extract the magnon coupling strength for the four-magnon scattering process (see Fig.~\ref{fig:paper_fig_2}(B) for graphical definition of $k_p$, $k_s$, and $k_i$) directly from its spatial magnon growth profile, thereby demonstrating the first component of the cascade of magnon scattering processes.

We apply a dual-tone microwave drive to the YIG film consisting of a pump at frequency $f_p$ and a signal $f_s$ that acts as a seed. In the perpendicular pumping configuration (see Supplementary Information), the dominant nonlinearity arises from the dipole-dipole interaction, which mixes these modes to generate an idler magnon at $f_i = 2f_p - f_s$. By tuning $f_p$ and $f_s$ such that the resulting idler frequency $f_i$ coincides with the NV spin transition frequency $f_{\rm NV}$, we use the NV ensemble magnetometer within the confocal volume as a spectrally selective filter to image only the nonlinear scattering product, effectively blind to the pump and signal fields; see Fig.~\ref{fig:paper_fig_2}(B).

Figure~\ref{fig:paper_fig_2}(C) shows a map of the measured spectral landscape of this process. We fix the pump frequency $f_p$ and sweep the seed $f_s$, measuring the NV photoluminescence (PL) as a function of distance from the antenna, using the pulse sequence in Figure~\ref{fig:paper_fig_2}(A)(i). We observe a rich structure of resonances. Beyond the expected direct driving of the NV spin transition frequency (labeled (iv) in Fig.~\ref{fig:paper_fig_2}(C)), we detect distinct magnon mixing orders, labeled as follows in Fig.~\ref{fig:paper_fig_2}(C): (i) the primary four-magnon signal at $f_i = 2f_p - f_s$; (iii) the inverse process $2f_s - f_p$; and (ii) a set of higher-order processes at $f_i = (n+1)f_p - nf_s$, where $n$ is an integer indicating the order. The visibility of these higher orders confirms the high coherence of the nonlinear four-magnon interaction process, a prerequisite for the formation of stable pre-thermal states.

We focus on the first-order idler mode (i) to extract the four-magnon interaction strength. Since the scattering is phase-coherent, the generated idler magnons drive coherent Rabi oscillations of the NV sensor. By measuring the Rabi frequency $\Omega_R$ as a function of position $y$ (Fig.~\ref{fig:paper_fig_2}(D)), using the pulse sequence in Fig.~\ref{fig:paper_fig_2}(A)(ii), we map the spatial envelope of nonlinear magnon generation. We convert this Rabi frequency to the local magnon population density $\langle n_{\bm k} \rangle$ via the relation (derived in the Supplementary Information):
\begin{equation}\label{eq:magnon_population}
    \langle n_{\bm k} \rangle = s_{\rm eff} \left(\frac{\Omega_R}{\gamma \mathcal{C}}\right)^2
\end{equation}
where $\gamma = \mu_0 \gamma_e$ is the electron gyromagnetic ratio scaled by the vacuum permeability, $s_{\rm eff} = M_s / (\gamma_e \hbar)$ is the effective spin density of the YIG film, set by the saturation magnetization $M_s = 1.55 \times 10^{5}$~A/m, and $\mathcal{C}$ collects the saturation magnetization together with the geometric, evanescent, and ellipticity factors.

To extract the four-magnon coupling $g^{pp}_{si}$, we fit the spatial growth of the idler population density to a coupled-mode theory derived from the nonlinear Schrödinger equation (see Supplementary Information). Assuming the pump and seed amplitudes remain undepleted over the measurement range, the idler amplitude $A_i(y)$ evolves as:
\begin{equation}\label{eq:idler_envelope}
    A_i(y) = \frac{A_0}{\lambda_0 - \lambda_i - i\Delta k^y} \left[ e^{-\lambda_i y} - e^{(-\lambda_0 + i\Delta k^y)y} \right]
\end{equation} 

Here, $A_0 = -i g^{pp}_{si} A_p^2 A_s^*/v^y_i$ is the nonlinear source term, $v^y_i$ is the $y$-component of the idler group velocity, $\lambda_{p,s,i}$ are the spatial decay rates of the pump, seed, and idler, $\lambda_0 = 2\lambda_p + \lambda_s$ is the combined decay rate of the source, and $\Delta k^y$ is the momentum mismatch ($\Delta k^y = 2k_p^y - k_s^y - k_i^y$) for the four-magnon scattering process. This model captures the competition between nonlinear gain and propagation loss, providing reasonable fits to the experimental data in magnon dynamics and spatial growth (Fig.~\ref{fig:paper_fig_2}(E)).

The extracted four-magnon interaction strengths $g^{pp}_{si}$ from this fitting procedure are plotted in Fig.~\ref{fig:paper_fig_2}(F) (purple dots), alongside theoretical predictions from standard T-matrix theory~\cite{krivosik_hamiltonian_2010} (orange dashed line). To calculate $g^{pp}_{si}$, we extract $A_0$ from the fits and determine $A_p$ and $A_s$ from the microwave power injected into the stripline, assuming a linear relationship between the drive current and the magnetization amplitude~\cite{bertelli_magnetic_2020}. We attribute the discrepancy between the measured and predicted values of $g^{pp}_{si}$ to our restriction to $k_x = 0$ in {} the scattering model. In practice, the large magnon populations injected at the pump and signal frequencies may also scatter into $k_x \neq 0$ modes at $f_{\rm NV}$, enhancing $\Omega_R$ and hence the inferred $A_0$, and thereby yielding a larger extracted $g^{pp}_{si}$ than predicted.

\subsection{Cascaded Parametric Instabilities}
\label{sec:cascaded}

Having characterized the elementary four-magnon vertex, we now turn to the regime of strong single-tone microwave driving near the ferromagnetic resonance (FMR) of the YIG film, where the system spontaneously organizes into a hierarchy of non-equilibrium states (Fig.~\ref{fig:paper_fig_3}(A)). Previous studies have hinted that high-power driving can redistribute magnon populations via three-magnon splitting~\cite{liu_time-resolved_2019} or four-magnon scattering~\cite{du_control_2017}.

We drive the system with a single frequency $\omega_0$ near the FMR and spatially scan the NV confocal volume to map the resulting magnon concentration. The measured NV ODMR spectral response, shown in Fig.~\ref{fig:paper_fig_3}(B) for a magnetic bias field of 22~mT, reveals a striking pattern. Beyond the direct excitation of the NV spin transition at 2.27~GHz, we observe a series of discrete, magnetic-field-dependent resonances (highlighted in red). These features signify that magnons driven near the FMR are undergoing spontaneous four-magnon scattering events, generating daughter populations that are resonant with the NV sensor. 

To describe this hierarchy of magnon interactions, we model the system as a cascade of parametric resonances. A pair of pump waves $(\omega_0, \mathbf{k}_0)$ scatters into a pair of daughter modes $(\omega_1, \mathbf{k}_1)$ and $(\omega_2, \mathbf{k}_2)$. Uniquely, our theoretical framework relaxes the standard on-shell assumption, and analyzes nonequilibrium dynamics in a system without translational invariance. By solving the coupled-mode equations in the steady state (see Supplementary Information), we derive the spatial growth rate $\lambda$ for daughter modes propagating perpendicular to the microwave stripline antenna (i.e., along the direction probed by the NV sensor, as shown by the white arrow in Fig.~\ref{fig:paper_fig_1}(B)):
\begin{equation}\label{eq:parametric_resonance}
    \lambda = \pm \sqrt{\frac{|g^{00}_{12} A_0^2|^2}{v^y_{\bm{k}_1} v^y_{\bm{k}_2}} - \left( \frac{\tilde{\omega}(\bm{k}_2)+ \tilde{\omega}(\bm{k}_1)- 2 \omega_0}{v_{\bm{k}_1}^y+ v_{\bm{k}_2}^y}\right)^2}
\end{equation}
Here, $g^{00}_{12}$ is the four-magnon interaction vertex, $A_0$ is the pump amplitude, $v^y_{\bm{k}_{1,2}}$ are the group velocities of the daughter modes, and $\tilde{\omega}(\mathbf{k})$ includes the mean-field frequency shifts. The structure of this equation reveals the dynamic selection mechanism of the cascade. The instability growth is maximized at the specific phase-space coordinates where the resonant manifold (the black surface satisfying the on-shell condition) intersects with the contours of vanishing group velocity ($v^y \rightarrow 0$, red lines in Fig.~\ref{fig:paper_fig_3}(A)).  The gain term in \equ{eq:parametric_resonance} scales as $1/\sqrt{v^y_{\bm{k}_1} v^y_{\bm{k}_2}}$ and is sharply enhanced as the daughter group velocities approach zero. As modes with low group velocities experience an extended spatial overlap with the active pumping region, they accumulate a larger amplitude per unit propagation length, maximizing the parametric gain along these slow-magnon lines.

\begin{figure*}[t]
    \centering
    \includegraphics[width=\textwidth]{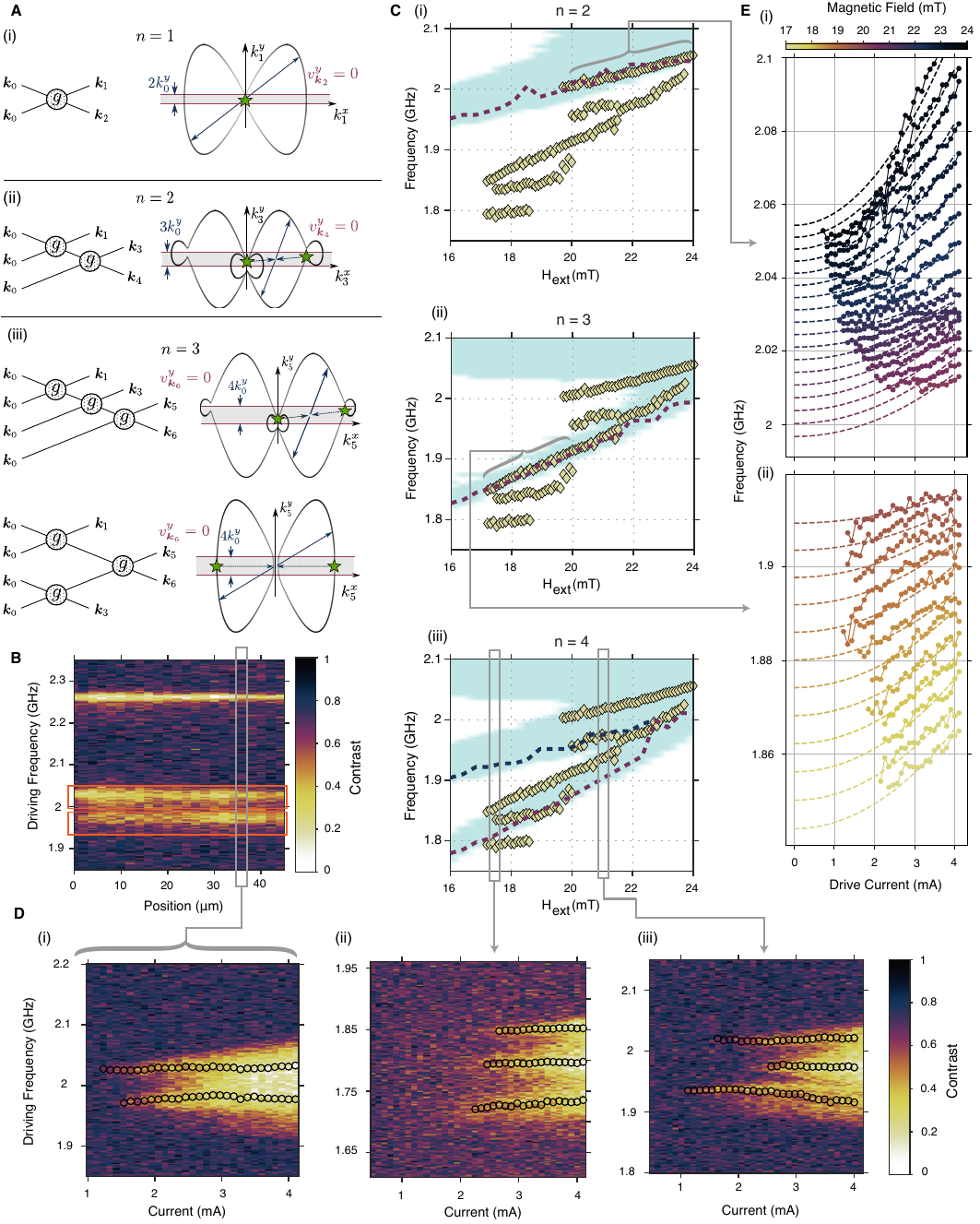}
\end{figure*}

\begin{figure*}[t]
    \caption{\textbf{(A)} Three successive stages of the cascaded magnon scattering process are illustrated. (i) A pair of pump magnons at $\bm k_0$, indicated by the stars in the right panel, scatters into daughter modes $\bm k_1$ and $\bm k_2$. The bow-tie-shaped contour represents the on-shell phase space available to these daughter modes. The red contours mark vanishing $y$-component of the group velocity and delimit the gray region in which exponential spatial growth is allowed. The strongest instability occurs near the intersection of the on-shell contour with a slow-magnon line, and the corresponding daughter mode becomes a pump for the next generation. (ii) and (iii) show the subsequent scattering stages, in which the most unstable mode from the preceding generation acts as a new pump, producing modified on-shell phase-space contours and slow-magnon lines. \textbf{(B)} Example measured NV ODMR spectrum along the diamond chip and hence the YIG film, as a function of scan position and single-tone microwave drive frequency at fixed bias magnetic field (22\,mT). The zero to one scale of this plot indicates normalized ODMR photoluminescence (PL) contrast (see Supplementary Information). Discrete NV PL resonances (highlighted in red) are observed when cascaded four-magnon scattering populates magnons at the NV frequency. \textbf{(C)} Comparison of each identified spin transition resonance frequency (diamonds), identified from a series of measured ODMR spectra, plotted as a function of bias magnetic field at a fixed position approximately 3--5\,$\upmu$m from the microwave stripline antenna, overlaid with expected PL resonance frequencies (dashed lines) for each magnon cascade order calculated from the two-dimensional model (see Supplementary Information). We find reasonable agreement between the predicted and measured (i)-(iii) resonances up to fourth order. Blue shading indicates the regions in which the model permits parametric instability, while the dashed lines mark the frequencies of maximum calculated growth rate. \textbf{(D)} NV ODMR spectra, as a function of microwave drive current, around the magnon-scattering-induced PL resonances in (B) and (C); (i)-(iii) are measured at fixed spatial positions and bias magnetic fields (indicated by gray boxes) in (B) and (C). At each driving current, a peak finding algorithm is used to extract the central PL frequency of each resonance as a function of microwave drive current (circular markers). \textbf{(E)} Center frequencies of two sets of magnon-scattering-induced PL resonances from (C) plotted as a function of drive current through the microwave stripline antenna. Each set of points (PL resonance frequencies) represents a different bias magnetic field, as given in the legend above. Fits to our quadratic model given in \equ{eq:cascade_resonance} are plotted as dashed lines for each bias field.}
    \label{fig:paper_fig_3}
\end{figure*}

Once a daughter magnon population reaches a critical threshold, it acts as a secondary pump for a subsequent magnon scattering generation. This stepwise thermalization explains the discrete bands observed in our data: each NV PL dip corresponds to a specific generation ($n$) of the cascade. We track these resonances across a broad magnetic field range (17--24~mT, Fig.~\ref{fig:paper_fig_3}(C)), confirming their robust nature.

However, the mode with the highest theoretical growth rate is not necessarily the dominant signal in our measurement. As detailed in the Supplementary Information~\cite{supplement}, the NV sensor acts as a momentum filter, sensitive only to modes within a finite window of $k$-space defined by the sensor-sample standoff distance. While the global instability maximum often occurs at large transverse wavevectors inaccessible to our sensor, the observed NV PL signal in Fig.~\ref{fig:paper_fig_3} arises from integrating the instability spectrum over the accessible momentum window. Consequently, off-resonant contributions, where parametric gain dominates the off-shell penalty in \equ{eq:parametric_resonance}, contribute significantly to the integrated signal. This effect explains the robustness of the observed bands: the NV sensor captures the spectral tails of the instability, allowing for the detection of discrete scattering patterns even when the divergence peak lies outside the maximal sensitivity window.

Having established that these discrete resonances are robustly detected across a range of magnetic fields (Fig.~\ref{fig:paper_fig_3}(C)), we next examine their evolution with microwave drive power. We measure the NV PL spectrum at a fixed spatial position while sweeping the drive current (Fig.~\ref{fig:paper_fig_3}(D)), and extract the central frequency of each resonance at each power level. A signature feature revealed by this measurement is the microwave-power-dependence of the observed resonance frequencies. As shown in Fig.~\ref{fig:paper_fig_3}(E), the resonance positions shift quadratically with microwave drive current. This effect arises from the Hartree-Fock mean-field shift $\omega_{\rm NL} \propto \beta I^2$ induced by the high magnon density. As the driven magnon amplitude scales linearly with current ($A_0 \propto I$), the population-dependent mean-field shift scales quadratically ($\langle n_{\bm k} \rangle \propto |A_0|^2 \propto I^2$) (see Supplementary Information). For an $n$-th order cascade, the resonance condition becomes:
\begin{equation}\label{eq:cascade_resonance}
    \omega_0(n, I) = \frac{\omega_{\rm NV} + (n-1)\Delta_0}{n} + \frac{2(n-1)}{n} \beta I^2
\end{equation}
where $\Delta_0$ is the FMR gap. We fit the power-dependent resonance spectrum to this model (dashed lines in Fig.~\ref{fig:paper_fig_3}(E)), allowing us to extract the cascade order $n$ and the effective nonlinearity coefficient $\beta$.

Taken together, the two-dimensional resonance model and the power-dependent fits reveal distinct families of instabilities. The higher-field branch (Fig.~\ref{fig:paper_fig_3}(C)(i) and (E)(i)) is consistent with an $n \approx 2$ cascade process, while the lower-field branch (Fig.~\ref{fig:paper_fig_3}(C)(ii) and (E)(ii)) corresponds to $n \approx 3$. A third, weaker branch (Fig.~\ref{fig:paper_fig_3}(C)(iii)) is consistent with an $n \approx 4$ cascade in the two-dimensional model, demonstrating that the hierarchy extends to at least four generations of daughter magnons before the signal falls below our detection threshold. The deviation from integer values extracted from the power-dependent fits indicates that the dominant scattering channels involve modes with non-zero transverse momentum ($k_x \neq 0$). This result is physically expected, as the most unstable modes reside at finite $k_x$, where the system can simultaneously satisfy the vanishing group velocity condition and minimize the off-shell penalty. This mode mixing is further evidenced by the positive sign of the extracted $\beta$ coefficients (see Supplementary Information), a feature that T-matrix theory predicts only for modes with finite $k_x$.

\section{Discussion}\label{sec:discussion}

Our results provide a spatially resolved map of the specific kinetic pathways through which a driven bosonic fluid redistributes energy away from equilibrium. By deconstructing the approach to equilibrium in a YIG thin film, we reveal that the pre-thermal regime is not a featureless, diffusive continuum, but a structured hierarchy of discrete interactions. The observation of these cascaded parametric instabilities challenges the standard continuum assumptions of Wave Turbulence Theory~\cite{zakharov1992kolmogorov, nazarenko2011wave}. Instead of a smooth flow of energy toward the thermal bath, we observe a cascade of parametric magnon scattering events, where energy accumulates along specific slow-magnon contours in phase space.

This discrete anatomy implies that the pathway to equilibrium is governed by strict dynamic selection rules. We find that the driven YIG film naturally selects scattering channels that balance two competing forces: the maximization of parametric gain (favored by vanishing group velocity) and the minimization of the off-shell penalty. This behavior mirrors the ``fireworks'' emission in driven Bose--Einstein condensates~\cite{clark_collective_2017, fu_density_2018}, suggesting a universality in how driven systems restructure themselves near kinetic bottlenecks.

These observations are enabled by the strong four-magnon nonlinearity and low intrinsic damping of YIG, which together allow the parametric gain to overcome dissipation and sustain several generations of scattering at accessible microwave drive powers. Crucially, our stimulated wave-mixing measurement provides direct, spatially resolved access to the underlying four-magnon vertex $g$, rather than relying on an analytically approximated coupling~\cite{krivosik_hamiltonian_2010}. Anchoring the cascade model in this measured interaction is what allows us to quantitatively reproduce the observed hierarchy of resonances and to establish the dynamic selection rules as the organizing principle of the pre-thermal cascade.

Methodologically, our work establishes spatially resolved NV magnetometry as a powerful technique for ``chronocounting'' quasi-particle interactions. By using propagation distance as a proxy for interaction time, we transformed a steady-state measurement into a dynamical map of the scattering evolution. Unlike inductive or optical techniques that integrate over large volumes, the sub-micron resolution of the NV sensor allows us to resolve the fine structure of these instabilities and directly extract the microscopic interaction strength local to the scattering events.

Looking forward, this platform offers a controlled measurement testbed for simulating the nonlinear kinetics of diverse physical systems. The ability to tune the interaction vertex via geometry and the cascade hierarchy via drive power opens the door to studying the crossover from discrete parametric cascades to hydrodynamic flow. Future experiments could probe the phase-space correlations of these scattering events. Because parametric amplification generates correlated magnon pairs, the daughter populations should retain distinct statistical signatures absent in a purely stochastic wave-turbulent cascade. Spatially mapping the second-order coherence function $g^{(2)}$ would directly expose these pairwise correlations, providing a definitive test of the discrete relaxation pathways. By bridging the gap between microscopic scattering rules and macroscopic fluid dynamics, thin-film magnonics provides a tangible window into the non-equilibrium evolution of quantum many-body systems.

\section{Methods}\label{sec:methods}

\subsection{Sample Details}

We use a 4\,mm $\times$ 6\,mm sample chip diced from a 100\,nm-thick YIG thin film grown on a 3-inch-diameter GGG substrate. The sample chip is patterned with a stripline antenna, with a width of 10\,$\upmu$m made from a 10\,nm Ti, 300\,nm Au stack, with wide pads on either end. The chip is mounted on a PCB, with both ends of the stripline antenna wirebonded to PCB traces: one side connects to the microwave drive circuit, and the other to a 50~$\Omega$ termination. The circuit drives currents at selected frequencies through the stripline antenna.

On the YIG film, we place a 50\,$\upmu$m by 50\,$\upmu$m diamond chip. This chip, with thickness ${\sim}1$--$2$~$\upmu$m, was commercially fabricated from a bulk single-crystal CVD diamond substrate (Element Six, natural isotopic abundance) cut along the $\langle 111 \rangle$ direction. The diamond is implanted with nitrogen ions with implantation energy of 10\,keV and a dose calibrated to yield $>1000$ NVs per $\upmu$m$^2$, without extra steps to create additional vacancies and anneal the diamond~\cite{asif_diamond_2024}. The resulting implanted nitrogen atoms are at a depth of approximately 10\,nm from the diamond surface~\cite{asif_diamond_2024}, with approximately 50 NVs per confocal volume along each of the four NV orientation classes within the diamond~\cite{barry_sensitivity_2020}. On the surface of the diamond chip with the NV centers, we deposit a 210\,nm-thick layer of TiO$_2$ via ALD to increase the separation between the NVs and the YIG surface. We then use a pick and place technique, using a probe station and tungsten tips, to place the diamond chip on the YIG, with the NV side facing (in proximity to) the YIG surface. The diamond chip is placed approximately 3--5\,$\upmu$m from the antenna and is not perfectly flush with the YIG surface (see Fig.~\ref{fig:paper_fig_1}(B)). The NV/YIG separation across the diamond chip (in the direction of the white arrow in Fig.~\ref{fig:paper_fig_1}(B)) is measured using NV $T_1$ relaxometry measurements (see the Supplementary Information), and is between 250--500\,nm. Data is also collected on a second diamond chip to confirm the general behavior; but all data presented in this paper and related quantitative calculations are acquired using a single chip.

\subsection{Experimental Setup}

We utilize a confocal microscope with a 0.9\,NA objective and an estimated 350\,nm spot size. It is equipped with a 520\,nm Cobolt diode laser to initialize the NVs into the $m_s = 0$ state, and a single photon detector to measure the NV photoluminescence (PL). The diamond/YIG sample is mounted on a 3-axis Newport translation stage. The focus of the objective is fixed in space and we sweep the stage position to scan across the diamond chip. To apply the bias field, we use a permanent magnet mounted on an independent translation stage. We calibrate the position of the magnet to specify the field magnitude and to reliably align the field along one of the four NV ensemble orientation classes within the diamond~\cite{barry_sensitivity_2020}, and we calculate the associated lab-frame directions of these classes. Due to the $\langle 111 \rangle$ orientation of the diamond, we apply the field $-71^\circ$ from the surface normal.

The microwave circuit is driven by two signal generators (Agilent E4428 and Rohde \& Schwarz SMA100A) and terminated with a Signal Hound spectrum analyzer, serving as the 50~$\Omega$ load. The signals are pulsed using Mini-Circuits switches. These are driven using a Tektronix AWG, which is programmed using our experimental control software. The AWG is also used to modulate the laser diode. NV PL measurements from the single photon detector are digitized using a National Instruments DAQ. The pulse sequence creation, data collection, and analysis are performed using a custom Matlab interface on our lab computer.

\section{Acknowledgments}
We gratefully acknowledge discussions with Nikola Maksimovic, Ruolan Xue, Shaowen Chen, and Elizabeth Park. This work was supported by the SNSF project 200021 212899, SNSF Sinergia grant CRSII--222792, Swiss State Secretariat for Education, Research and Innovation (contract number UeM019-1), NCCR SPIN, a National Centre of Competence in Research, funded by the Swiss National Science Foundation (grant number 225153). This work was supported by the U.S. Army Research Laboratory, under Contract No. W911NF2420143; the Laboratory for Physical Sciences Jumping Electron Quantum Fellowship Program under Award No. H9823022C0029; and the University of Maryland Quantum Technology Center. A. Y. is partly supported by the Gordon and Betty Moore Foundation through Grant GBMF 12762, by the U.S. Army Research Office (ARO) MURI project under grant number W911NF-21-2-0147, and by the  Army Research Office (ARO) grant number W911NF-22-1-0248

\bibliography{paper_citations}

\end{document}


\title[Supplementary Information]{Supplementary Information for: Spatially Resolving the Pre-Thermal Anatomy of a Driven Bosonic Fluid}

\author[1,2]{\fnm{Shantam} \sur{Ravan}}
\author[3]{\fnm{Aaron} \sur{Müller}}
\author[1,4]{\fnm{Johannes} \sur{Cremer}}
\author[3]{\fnm{Jonathan} \sur{Curtis}}
\author[1]{\fnm{Daniel} \sur{Fernandez}}
\author[4]{\fnm{Ronald} \sur{Walsworth}}
\author[3]{\fnm{Eugene} \sur{Demler}}
\author[1]{\fnm{Amir} \sur{Yacoby}}

\affil[1]{\orgdiv{Department of Physics}, \orgname{Harvard University}, \orgaddress{12 Oxford Street, Cambridge, MA 02138, USA}}
\affil[2]{\orgdiv{Department of Physics}, \orgname{University of Maryland, College Park}, \orgaddress{4296 Stadium Dr, College Park, MD 20742, USA}}
\affil[3]{\orgdiv{Department of Physics}, \orgname{ETH Zurich}, \orgaddress{Otto-Stern-Weg 1, 8049 Zürich, Switzerland}}
\affil[4]{\orgdiv{Quantum Technology Center}, \orgname{University of Maryland, College Park}, \orgaddress{8228 Paint Branch Dr, College Park, MD 20742, USA}}

\maketitle

\section{Sample Fabrication and Experimental Setup}\label{sec:sample}

\subsection{YIG Thin Film}

As mentioned in the main text, we use a 4\,mm $\times$ 6\,mm chip diced from a 100\,nm thick YIG thin film grown via LPE on a GGG substrate. This is a commercially available 3 inch wafer purchased from Matesy GmbH. The characteristics of this film are similar to those used in~\cite{zhou_magnon_2021, bertelli_magnetic_2020}, including low damping and a sharp FMR linewidth. This is patterned with a stripline antenna, and an NV chip with an ALD coating is placed on the surface of the YIG adjacent to it.

We use the technique described in \cite{zhou_magnon_2021} to image magnon wavelength for a series of magnetic field values. Using this information, we fit the dispersion relation of the magnons to obtain the saturation magnetization of the material. For the other constants, we use the generally accepted values in the literature. The relevant YIG film parameters used throughout this work are:
\begin{align}
    D &= 100\,\text{nm}, && M_s = 1.55 \times 10^{5}\,\text{A/m} \;(= 155\,\text{emu/cc}), \notag \\
    \alpha &= 10^{-4} && A_{\text{ex}} = 3.7\times 10^{-12}\, \text{J/m} \; (=3.7 \times 10^{-7}\,\text{erg/cm}),  
    \label{eq:YIG_parameters}
\end{align}
where $D$ is the film thickness, $M_s$ is the saturation magnetization, $A_{\text{ex}}$ is the exchange stiffness, and $\alpha$ is the Gilbert damping constant. Additionally, the NV axis makes an angle $\theta_{\rm NV} = -71^\circ$ with the film normal.

\subsection{Confocal Microscope and Microwave System}

To measure our sample, the experiment has two main components: a confocal microscope and a microwave circuit. Expanding on the information presented in the main text, our confocal microscope uses a Cobolt 06-MLD diode laser at 520\,nm to initialize and read out the NV centers. The excitation and collection path are both directed through a Nikon 0.9\,NA objective and a 350\,nm spot size is estimated. The collection path is filtered twice, once through a dichroic beamsplitter and then through a longpass filter to remove any remaining laser light. The collected light is focused through a pinhole and then detected by a single photon detector, which directs counts to a NI PCI-6221 DAQ connected to a computer.

This computer is the nexus of the experiment. We use a custom built Matlab interface to control the experiment. This is used to read out the DAQ, program the two signal generators, and coordinate the pulse sequences via a Tektronix AWG5014C arbitrary waveform generator.

\section{NV Center Physics}\label{sec:nv-physics}

The NV center forms an electronic spin triplet, where the $m_S = \pm 1$ states are degenerate and split from the $m_S = 0$ state by the zero-field splitting $\Delta/(2 \pi) = 2.87$ GHz. Applying a static magnetic field $H^z$ lifts the degeneracy of the $m_S = \pm 1$ states. The Hamiltonian of a single NV center located at $\bm r_i$ interacting with a transverse oscillating magnetic field is:
\begin{align}
    \mathcal H_{\rm NV} =  \frac{\Delta - \gamma H^z}{2}\left[\sigma^0-\sigma^z\right]+ \frac{\gamma}{\sqrt{2}}\left[H^+(\bm r_i, t)\sigma^- +H^-(\bm r_i, t) \sigma^+\right]
\end{align}
where $\sigma^0$ is the identity matrix, $\sigma^\pm = (\sigma^x\pm i \sigma^y)/2$, and $H^\pm = H^x\pm iH^y$~\cite{rustagi_sensing_2020}. Throughout, $\gamma = \mu_0 \gamma_e$ denotes the electron gyromagnetic ratio scaled by the vacuum permeability, with $\gamma_e = 2\pi \times 28$\,GHz/T, so that $\gamma H$ is an angular frequency for fields and magnetizations expressed in A/m.

\subsection{Coherent Dynamics: Rabi Frequency}

When the local transverse field is a monochromatic coherent wave on resonance with the lower transition ($\omega \approx \Delta - \gamma H^z$), the system undergoes deterministic Rabi oscillations between $\ket{0}$ and $\ket{-1}$. Evaluating the Schrödinger equation in the rotating wave approximation yields a local transition probability $P_i(t) = \sin^2(\Omega_{R,i} t)$, where the squared local Rabi frequency is defined directly by:
\begin{align}
    \Omega_{R,i}^2 = \frac{\gamma^2}{2} H^+(\bm r_i, t) H^-(\bm r_i, t) \,.
    \label{eq:local_rabi_squared}
\end{align}

\subsection{Incoherent Dynamics: $T_1$ Transition Probability}

Conversely, evaluating the transition probability via Fermi's Golden Rule allows us to establish the incoherent NV spin $T_1$ dynamics on the same footing. Treating the transverse field as a perturbation $\mathcal H'$ in the interaction picture, the probability of an upward transition $P_i^\uparrow \equiv P^{0\leftarrow-1}_i$ evaluates to:
\begin{align}
    P_i^\uparrow &= \left|\frac{1}{i}\int_0^t\d t' \bra{0} \mathcal H'_I(t')\ket{-1}\right|^2 \notag \\
    &= \frac{\gamma^2}{2} \int_0^t \d t_1 \d t_2 e^{-i \omega_{\rm NV}(t_1-t_2)} H^-(\bm r_{\rm NV}^i, t_1) H^+(\bm r_{\rm NV}^i, t_2) 
\end{align}
where $\omega_{\rm NV} = \Delta - \gamma H^z$. Similarly, for the downward process $P_i^\downarrow \equiv P^{-1\leftarrow0}_i$:
\begin{align}
     P_i^\downarrow  = \frac{\gamma^2}{2} \int_0^t \d t_1 \d t_2 e^{i \omega_{\rm NV}(t_1-t_2)} H^+(\bm r_{\rm NV}^i, t_1) H^-(\bm r_{\rm NV}^i, t_2) 
\end{align}
Taking Fourier transformations isolates the frequency filtering components:
\begin{align}
    P_i^{\uparrow, \downarrow} &= \frac{\gamma^2}{2} \int \frac{\d \omega \d \omega'}{(2 \pi)^2} W_{\rm T1}^{\uparrow, \downarrow} (\omega, \omega')  K_i^{\uparrow, \downarrow}(\omega, \omega') \label{eq:transition_prob}
\end{align}
where we introduced the magnetic field kernels:
\begin{align}
      K_i^{\uparrow}(\omega,\omega')&=\int \d t_1 \d t_2  e^{-i \omega t_1 - i \omega' t_2} H^-(\bm r_{\rm NV}^i, t_1) H^+(\bm r_{\rm NV}^i, t_2) \, , \\
      K_i^{\downarrow}(\omega,\omega')&=\int \d t_1 \d t_2  e^{-i \omega t_1 - i \omega' t_2} H^+(\bm r_{\rm NV}^i, t_1) H^-(\bm r_{\rm NV}^i, t_2) \, ,
\end{align}
and the corresponding frequency filtering functions:
\begin{align}
    W_{\rm T1}^{\uparrow} (\omega, \omega') &= \frac{e^{-i(\omega_{\rm NV} -\omega) t}-1}{(\omega_{\rm NV} -\omega)} \frac{e^{i(\omega_{\rm NV} +\omega') t}-1}{(\omega_{\rm NV} +\omega')} \, , \\
    W_{\rm T1}^{\downarrow} (\omega, \omega') &= \frac{e^{i(\omega_{\rm NV} +\omega) t}-1}{(\omega_{\rm NV} +\omega)} \frac{e^{-i(\omega_{\rm NV} -\omega') t}-1}{(\omega_{\rm NV} -\omega')} \, . \label{eq:NV_freq_filter}
\end{align}
For large enough times $t\gg \tau$ (where $\tau$ is the decay timescale of $K_i^{\uparrow,\downarrow}$), $W_{\rm T1}^{\uparrow}(\omega, \omega')$ is strongly peaked at $\omega \approx \omega_{\rm NV}$ and $\omega' \approx -\omega_{\rm NV}$, in which case we find $W_{\rm T1}^{\uparrow}( \omega_{\rm NV}, -\omega_{\rm NV}) = t^2$. The same holds for interchanging $\omega$ and $\omega'$ in the downward process.

\section{Magnon Physics and NV Coupling}\label{sec:magnon-interaction}

We present the effective Hamiltonian of the driven magnon system, assuming a quartic nonlinearity. The third-order nonlinearity is kinematically disallowed in our target frequency range. The effective Hamiltonian is:
\begin{equation}
    \frac{\mathcal{H}}{V} = \sum_{\bm k} (\omega_{\bm k} - i \Gamma_{\bm k}) \Psi_{\bm k}^* \Psi_{\bm k} + \frac{1}{2} \sum_{\bm k_1, \bm k_2, \bm k_3, \bm k_4} g_{34}^{12} \delta_{34}^{12} \Psi_{\bm k_1}^* \Psi_{\bm k_2}^* \Psi_{\bm k_3} \Psi_{\bm k_4} + \sum_{\bm k} \left[S_{\bm k}(t) \Psi_{\bm k}^*(t) + \text{c.c.} \right]
    \label{eq:Hamiltonian_momentum}
\end{equation}
The first term captures the kinetic energy of magnons with dispersion relation $\omega_{\bm k}$ and Gilbert damping  $\Gamma_{\bm k}=\alpha \omega_{\bm k}$. The second term describes momentum-dependent scattering with coupling strength $g_{34}^{12}$, where $\delta_{34}^{12}$ enforces momentum conservation. The final term models the coherent stripline drive $S_{\bm k}(t)$. This Hamiltonian yields the equations of motion: 
\begin{align}
    i \partial_t \Psi_{\bm k}(t) &= \frac{\delta \mathcal H/V}{\delta \Psi^*_{\bm k}(t)} =  (  \omega_{\bm k}  -  i \Gamma_{\bm k} )  \Psi_{\bm k} + \sum_{\bm k_1, \bm k_2 ,\bm k_3} g_{3k}^{12} \delta_{3k}^{12} \Psi_1 \Psi_2 \Psi_3^* + S_{\bm k}(t)  \label{eq:NLWE_momentum} 
\end{align}
Before solving this equation, we connect the phenomenological constants $\omega_{\bm k}$, $g_{34}^{12}$, and $S_{\bm k}$ to their microscopic origins and demonstrate how this continuum model governs the interaction with the macroscopic NV ensemble.

\subsection{Dispersion Relation}

The derivation of the dispersion relation $\omega_{\bm k}$ of exchange dipole magnons in thin films is established in standard literature \cite{rustagi_sensing_2020, kalinikos_theory_1986}. The full dispersion relation evaluates to $\omega(k, \phi_k) = \sqrt{\omega_2 \omega_3 - \omega_1^2}$, where:
\begin{align}
    \omega_1 &= \gamma H_{\rm d} \sin\phi_k \cos\phi_k \cos\theta_0 f_k, \\
    \omega_2 &= \gamma \left[ H_{\rm int} + H_{ \rm ex} k^2 + H_{\rm d} \cos^2\phi_k f_k \cos^2\theta_0 + H_{\rm d} (1 - f_k) \sin^2\theta_0 \right], \\
    \omega_3 &= \gamma \left[ H_{\rm int} + H_{ \rm ex} k^2 + H_{\rm d} \sin^2\phi_k f_k \right],
\end{align}
where the demagnetization field is $H_{\rm d}=M_s$ and the exchange field is $ H_{ \rm ex}=2A_{\rm ex}/(\mu_0 M_s)$. The internal magnetic field is $H_{\rm int}= H_{\text{ext}} \cos(\theta-\theta_0) - H_{\text{d}} \cos^2(\theta_0)$ where $\theta$ specifies the angle of the applied external magnetic field $H_{\rm ext}$, and $\theta_0$ represents the equilibrium magnetization angle. The demagnetization form factor is $f_k = 1 - (1 - \exp(- k D))/(k D)$.

Given our magnetic field aligns perpendicular to the axis of magnon propagation ($\phi_k = \pi/2$), we evaluate the Damon--Eshbach (DE) mode~\cite{damon_magnetostatic_1961}:
\begin{align}
    \omega_{\text{DE}}(k) = \gamma \sqrt{\left[H_{\text{ext}} + H_{ \rm ex} k^2 + H_{\rm d} (1-f_k) \right] \left[H_{\text{ext}} + H_{ \rm ex} k^2 + H_{\rm d} f_k \right]} \, . \label{eq:dispersion_relation_full_DE}
\end{align}
Expanding this to linear order yields $\omega_{\rm lin}^{\rm DE}(k_y) = \Delta_0 + v_y k_y$, where the gap is $\Delta_0 = \gamma \sqrt{H_\text{ext} (H_{\rm d} + H_\text{ext})}$ and the group velocity is $v_y \equiv v_{\rm DE} = \gamma H_{\rm d}^2 D/(4 \sqrt{H_\text{ext} (H_{\rm d} + H_\text{ext})})$. This linear form underlies the analytic selection rules and resonance condition derived in \secu{sec:cascaded}; the numerical idler and growth-rate calculations (\secu{sec:stimulated}, \secu{sec:arnold_tongues}) instead use the exact group velocity $v_g = \partial \omega/\partial k$, which falls below $v_{\rm DE}$ where the dispersion flattens at the experimental wavevectors.

\subsection{Injection Profile}

The magnetic field generated by the stripline determines the injection profile $S_{\bm k}$. Evaluating the Biot-Savart law for a current density $\bm J(\bm\rho,z, t) = \hat x \, I/(WH) f(y/W) \cos(\omega_d t)$ localized at $z=D/2$ yields the in-plane Fourier transformed field:
\begin{align}
    \bm B_{\bm k}(t) = -\mu_0 I \delta(k_x) h(k_y H) w(k_y W)  \, [\hat y + i \, \text{sgn}(k_y) \hat z ]\cos(\omega_d t)
\end{align}
where $w(k_y W)$ is the Fourier transform of the lateral current distribution and $h(k_y H) = (1-\exp(- |k_y| H))/(|k_y| H)$. 

The stripline field couples linearly to the dynamic magnetization. After projection onto the driven Damon--Eshbach branch, the effective source inherits the geometric momentum profile and directional polarization of the microwave field. We therefore parameterize it as
\begin{align}
    S_{\bm k}(t) = i \, S_0 \, h(k_y H) w(k_y W) \theta(k_y) \delta_{k_x} \cos(\omega_d t)
\end{align}
where $S_0\propto I$ contains the polarization and mode-overlap factors. The form factors $h$ and $w$ determine the momentum width of the injected mode, while $\theta(k_y)$ captures the chirality of the DE-modes, confirming that injection occurs exclusively at positive momenta for the top surface of the sample.

\subsection{Magnetic Field at the NV Center}

The stray magnetic field at the NV center generated by the underlying magnons is evaluated within the near-field approximation. Performing an in-plane Fourier transform maps the sample magnetization to the local NV position $\bm r_{\rm NV}^i$ via the effective magnetostatic dipole tensor $\tilde D_{\bm q}$:
\begin{align}
    H^\alpha(\bm \rho_{\rm NV}^i, t) &= \int \frac{\d \bm q}{(2 \pi)^2} e^{-i \bm q \cdot \bm \rho_{\rm NV}^i} \tilde D_{\bm q}^{\alpha \beta} \tilde m^{\beta}_{\bm q}(t) \, ,
\end{align}
where $\tilde D_{\bm q} = R_y (\theta_{\rm NV}) D_{\bm q}(z_{\rm NV}) R^T_y(\theta_0)$ applies the rotational matrices necessary to align the lab frame with the local quantization axes. The dipole tensor evaluates to:
\begin{align}
    D_{\bm q}(z_{\rm NV}) = -  e^{-q z_{\rm NV}} \frac{1- e^{-qD}}{2q} 
    \begin{bmatrix}
        q_x^2/q & q_x q_y/q & iq_x\\
        q_x q_y/q & q_y^2/q & iq_y\\
        iq_x & iq_y & - q
    \end{bmatrix} \, .
    \label{eq:Dq_def}
\end{align}

\subsection{Ensemble Averaging and Momentum Filter Tensors}

The optical measurement integrates photon counts across the confocal volume, mapping local NV spin transitions to a macroscopic extensive observable. The coherent transition rate depends directly on the spatial sum of the field products across the $i$ NV centers. Expanding this product into momentum space yields:
\begin{align}
    \sum_{i \in W} &H^\alpha(\bm \rho_{\rm NV}^i, t_1) H^{\alpha'}(\bm \rho_{\rm NV}^i, t_2)  \notag \\
    &=\int \frac{\d \bm q}{(2 \pi)^2} \int \frac{\d \bm q'}{(2 \pi)^2} \sum_i  e^{-i (\bm q +\bm q') \cdot \bm \rho_{\rm NV}^i} \tilde D_{\bm q}^{\alpha \beta} \tilde m^{\beta}_{\bm q}(t_1) \tilde D_{\bm q'}^{\alpha' \beta'} \tilde m^{\beta'}_{\bm q'}( t_2) \, .
\end{align}
Approximating the discrete spatial sum as a continuous integral over the NV density $1/d_{\rm NV}^2$ generates a momentum-conserving Dirac delta function for a sufficiently large confocal spot. This collapses the cross-momentum integration ($\bm q' = -\bm q$):
\begin{align}
    \sum_{i \in W} H^\alpha(\bm \rho_{\rm NV}^i, t_1) H^{\alpha'}(\bm \rho_{\rm NV}^i, t_2) \approx \frac{1}{d_{\rm NV}^2} \int \frac{\d \bm q}{(2 \pi)^2} \tilde D_{\bm q}^{\alpha \beta} \tilde D_{-\bm q}^{\alpha' \beta'}\tilde m^{\beta}_{\bm q}(t_1)  \tilde m^{\beta'}_{-\bm q}(t_2) \,.
\end{align}
Expressing the transverse magnetization components in the circular basis,
$\tilde m^\pm=\tilde m^x\pm i\tilde m^y$, organizes the tensor contraction into the geometric kernels $\tilde{\mathcal D}_{\bm q}^{\uparrow\downarrow,\sigma\sigma'}$. The superscripts $\uparrow$ and $\downarrow$ label the field combinations $H^-H^+$ and $H^+H^-$ associated with the two NV transition channels. These filter functions evaluate to:
\begin{align}
    \tilde{\mathcal{D}}^{\uparrow, \pm \pm}_{\bm q} &= \left[1- e^{-qD}\right]^2 \frac{ e^{-2 q z_{\rm NV}}}{16q^4}  \left[ -\left(q_y \pm i q_x \cos \theta_0 \right)^2 + q^2 \sin^2 \theta_0 \right] \notag\\
    &\qquad \qquad \qquad \qquad \qquad \qquad \qquad \times  \left[ q_x^2 \cos^2 \theta_{\rm NV} + \left( q_y + q \sin \theta_{\rm NV} \right)^2 \right]  \, , \notag \\
    \tilde{\mathcal{D}}^{\uparrow, \pm \mp}_{\bm q} &=  \left[1- e^{-qD}\right]^2 \frac{ e^{-2 q z_{\rm NV}}}{16q^4} \left[ q_x^2 \cos^2 \theta_0 + \left( q_y \mp q \sin \theta_0 \right)^2 \right] \notag\\
    &\qquad \qquad \qquad \qquad \qquad \qquad \qquad \times  \left[ q_x^2 \cos^2 \theta_{\rm NV} + \left( q_y + q \sin \theta_{\rm NV} \right)^2 \right]  \,, \notag \\
    \tilde{\mathcal{D}}^{\downarrow, \pm \pm}_{\bm q} &=  \left[1- e^{-qD}\right]^2 \frac{ e^{-2 q z_{\rm NV}}}{16q^4} \left[ -\left(q_y \pm i q_x \cos \theta_0 \right)^2 + q^2 \sin^2 \theta_0 \right] \notag\\
    &\qquad \qquad \qquad \qquad \qquad \qquad \qquad \times  \left[ q_x^2 \cos^2 \theta_{\rm NV} + \left( q_y - q \sin \theta_{\rm NV} \right)^2 \right]\, , \notag \\
    \tilde{\mathcal{D}}^{\downarrow, \pm \mp}_{\bm q}& = \left[1- e^{-qD}\right]^2 \frac{ e^{-2 q z_{\rm NV}}}{16q^4}  \left[ q_x^2 \cos^2 \theta_0 + \left( q_y \mp q \sin \theta_0 \right)^2 \right]\notag\\
    &\qquad \qquad \qquad \qquad \qquad \qquad\qquad  \times   \left[ q_x^2 \cos^2 \theta_{\rm NV} + \left( q_y - q \sin \theta_{\rm NV} \right)^2 \right].
    \label{eq:mom_filter_T1}
\end{align}
These kernels encode the angular selection rules and the evanescent decay. As plotted in \figu{fig:height_map}(a), given the experimental angle $\theta_{\rm NV} = -71^\circ$ and the equilibrium magnetization angle $\theta_0$, the dominant contributions along the propagation axis $q_y$ arise exclusively from the density-like terms $\tilde{\mathcal{D}}^{\downarrow, +-}_{\bm q}$ and $\tilde{\mathcal{D}}^{\uparrow, -+}_{\bm q}$. The anomalous terms ($\pm \pm$) are geometrically suppressed.

\subsection{Macroscopic Rabi Frequency}

When a magnon mode acquires a macroscopic population under a strong coherent drive, the stochastic fluctuations condense into a deterministic classical field, driving coherent transitions. 

Because the optical measurement integrates photon counts across the entire confocal volume, the observed photoluminescence contrast reflects the ensemble-averaged transition probability of the defects within the spot. To ensure thermodynamic consistency at room temperature, the ferrimagnetic lattice is mapped to an effective continuum ferromagnet using the effective macroscopic spin density $s_{\rm eff}$ (defined directly by the measured saturation magnetization via $M_s = \gamma_e \hbar s_{\rm eff}$). The transformation is:
\begin{align}
    \tilde m^+_{\bm q}(t) = M_s \sqrt{\frac{2}{s_{\rm eff}}} \alpha_{\bm q}(t) \, , \qquad \text{and} \qquad \tilde m^-_{\bm q}(t) = M_s \sqrt{\frac{2}{s_{\rm eff}}} \alpha^*_{-\bm q}(t) \,.
\end{align}
In the coherent driven regime, the dynamics are governed by deterministic, monochromatic spin waves $\alpha_{\bm q}(t) = \alpha_{\bm q} e^{-i \omega_{\bm q} t}$. The extensive sum of the squared Rabi frequencies across the illuminated NV ensemble evaluates to:
\begin{align}
    \sum_i \Omega_{R,i}^2 = \frac{\gamma^2 M_s^2}{s_{\rm eff} d_{\rm NV}^2} \int \frac{\d \bm q}{(2 \pi)^2} \tilde{\mathcal{D}}^{\downarrow, +-}_{\bm q}  \langle n_{\bm q} \rangle \,,
    \label{eq:final_rabi_sum}
\end{align}
where $\langle n_{\bm q} \rangle = |\alpha_{\bm q}|^2$.

\subsection{1D Propagation and Population Extraction}

We model the driven magnon mode as a 1D wavepacket propagating along the $y$-direction. Because the spatial decay length of the magnon fluid is much larger than its carrier wavevector, we approximate the momentum spectrum by a Dirac delta function: $\langle n_{\bm q} \rangle = \langle \tilde n_0\rangle (2 \pi)^2 A_{\rm spot}  \delta(q_x) \delta(q_y - q_0)$. Furthermore, because a plane wave generates a spatially uniform field magnitude, every defect in the ensemble experiences an identical driving amplitude. The extensive NV ensemble sum thus reduces to the local single-spin value scaled by the spot area: $d_{\rm NV}^2 \sum_i \Omega_{R,i}^2 \equiv A_{\rm spot} \Omega_R^2$.

We now evaluate the detection efficiency for the driven mode at $q_x = 0$. Shape anisotropy renders the precession elliptical: the mode carries an in-plane transverse component and a smaller out-of-plane component in quadrature, with amplitude ratio $\eta \le 1$ set by the dispersion~\cite{bertelli_magnetic_2020}. Both components contribute coherently to the stray field above the film, yielding the phase-dependent geometric factor:
\begin{align}
    \mathcal{G}_\eta(s, \theta_{\rm NV}, \theta_0) = \frac{1}{\sqrt{\eta}} \left(1 - s\, \eta \cos\theta_0\right) \left(s - \sin\theta_{\rm NV}\right) \, ,
    \label{eq:G_eta_def}
\end{align}
where $s = \operatorname{sgn}(q_y) = \pm 1$ encodes the propagation direction of the detected mode and $\theta_0$ is the small out-of-plane cant of the static magnetization. The factor $(1 - s\,\eta\cos\theta_0)$ sets the chirality of the spin-wave stray field, the factor $(s - \sin\theta_{\rm NV})$ projects the circularly polarized stray field onto the co-rotating component at the NV, and the prefactor $1/\sqrt{\eta}$ captures the larger transverse magnetization per magnon of an elliptical mode. The corresponding generalized filter tensor takes the form:
\begin{align}
    \tilde{\mathcal{D}}^{\rm eff}_\eta(q_0) = e^{-2|q_0|z_{\rm NV}} \left[\frac{1-e^{-|q_0|D}}{4}\right]^2 |\mathcal{G}_\eta|^2 \,.
\end{align}
Substituting these spatial boundaries and elliptical limits into \equ{eq:final_rabi_sum} cancels the spot area coefficients identically, yielding:
\begin{align}
    \Omega_R^2 = \frac{\gamma^2 M_s^2}{16s_{\rm eff}} e^{-2|q_0|z_{\rm NV}} \left[1-e^{-|q_0|D}\right]^2 |\mathcal{G}_\eta|^2 \langle \tilde n_{0} \rangle \,.
    \label{eq:Rabi_vs_n0}
\end{align}
Inverting this relation extracts the volumetric magnon population density $\langle \tilde n_0 \rangle$ directly from the measured Rabi frequency:
\begin{align}
    \langle \tilde n_0 \rangle = s_{\rm eff} \left(\frac{4\Omega_R}{\gamma\, M_s\, |\mathcal{G}_\eta|}\right)^2  \frac{e^{2|q_0| z_{\rm NV}}}{\left(1 - e^{-|q_0| D}\right)^2} \, .
    \label{eq:nq_from_Rabi_combined}
\end{align}
This matches the expression quoted as Eq.~(1) of the main text by setting $\bm q_0 \equiv \bm k$, with the coupling constant $\mathcal{C}$ collecting the saturation magnetization together with the geometric, evanescent, and ellipticity prefactors:
\begin{align}
    \mathcal{C} \equiv M_s\, e^{-|q_0| z_{\rm NV}}\frac{1 - e^{-|q_0| D}}{4}|\mathcal{G}_\eta| \, .
    \label{eq:C_definition}
\end{align}

\section{Height Mapping}\label{sec:height_mapping}

To perform the density mapping detailed above, we must determine the average separation between the NV ensemble and the YIG film ($z_{\rm NV}$). We exploit the $T_1$ measurement protocol, noting that the NVs detect magnetic noise within a specific momentum window governed by the $e^{-2qz_{\rm NV}}$ filter tensor suppression. We sweep the magnetic field from approximately 12\,mT to 30\,mT. As the applied field increases, the resonant wavevector $\bm k$ decreases, increasing sensitivity to thermal magnons until the relaxation rate peaks near $|\bm k| \approx 1/(2 z_{\rm NV})$. We fit this response to extract the local separation.

\subsection{Thermal Magnon Noise and $T_1$ Relaxation}

Unlike the Rabi response driven by a coherent mode, the incoherent $T_1$ transition rate is driven by broadband thermal magnetic field fluctuations. However, the geometric coupling remains strictly governed by the exact same momentum filter functions $\tilde{\mathcal{D}}^{\uparrow \downarrow,\sigma \sigma'}_{\bm q}$ established in \secu{sec:magnon-interaction}.

To model this thermal noise at room temperature and GHz frequencies ($\hbar\omega \ll k_BT$), we relate the equilibrium transverse magnetization fluctuations to the dynamic susceptibility $\chi_{\alpha\beta}(\bm q,\omega)$ via the classical fluctuation-dissipation theorem: 
\begin{align}
    C_{\alpha\beta}(\bm q,\omega) = \frac{2k_BT}{\omega}\,\mathrm{Im}\,\chi_{\alpha\beta}(\bm q,\omega) \,.
\end{align}
Substituting this thermal spectrum into the momentum integrals established by Fermi's Golden Rule yields the explicit functional form for $\Gamma^{\uparrow,\downarrow}(H_{\rm ext})$. {}

\subsection{Height Map Calculation}

\begin{figure*}[t]
    \centering
    \includegraphics[width=\linewidth]{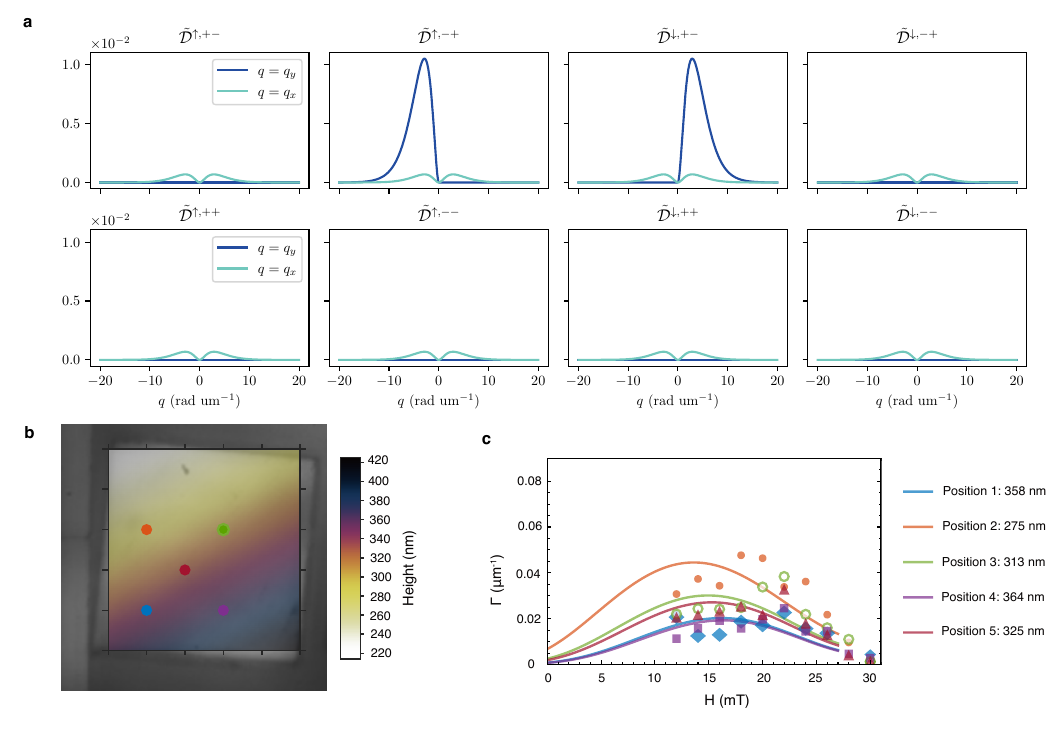}
    \caption{
        \textbf{Height mapping of NV-YIG separation.}
        (a) Momentum filter function $\tilde{\mathcal{D}}^{\downarrow,+-}_{\bm q}$ for $\theta_{\rm NV} =-71^\circ$ and the equilibrium magnetization angle $\theta_0$. Due to the geometry, the dominant contributions along the propagation axis $q_y$ arise from the density-like tensor elements.
        (b) Spatial map of the diamond chip showing the five measurement positions and the extracted height map of $z_{\rm NV}$ across the chip. The variation in $z_{\rm NV}$ (250--500~nm) reflects non-uniform contact between the diamond and YIG surfaces.
        (c) $T_1$ relaxation rate vs.\ applied magnetic field at five different positions on the diamond chip. Solid lines show fits to the thermal magnon noise model with $z_{\rm NV}$ as the free parameter.
    }
    \label{fig:height_map}
\end{figure*}

By performing the $T_1$ measurements across the diamond chip, we construct a spatial map of the NV-YIG separation. We plot these data points in Fig.~\ref{fig:height_map}(c), along with the corresponding fits to the thermal magnon noise model. The fitted $z_{\rm NV}$ values range from approximately 250~nm to 500~nm as seen in Fig.~\ref{fig:height_map}(b), reflecting non-uniform contact between the diamond and YIG surfaces. We use the locally-determined $z_{\rm NV}$ at each position to accurately calculate the magnon population densities from the measured Rabi frequencies.

\section{Stimulated Four-Magnon Scattering Model}\label{sec:stimulated}

In this section and the next, we derive the theoretical framework for understanding four-magnon scattering processes observed in our experiments. We first establish a general approximation scheme for solving the nonlinear wave equation, then apply it to the two-tone (stimulated scattering) and single-tone (cascaded scattering) experiments.

\subsection{Approximation Scheme for Driven Nonlinear Systems}

Our starting point is the nonlinear wave equation \equ{eq:NLWE_momentum} introduced in \secu{sec:magnon-interaction}. The full equation couples a continuum of momentum modes through the momentum-dependent four-magnon vertex, while also retaining propagation, damping, and the localized external drive. A direct time-domain solution would be computationally demanding and offer limited interpretability of the individual resonances observed experimentally. The measurements instead probe a weakly nonlinear, narrowband, continuous-wave regime in which the dominant scattering pathways can be organized systematically around the directly driven mode. We use the following experimentally motivated approximations:
{}

\begin{enumerate}
    \item \textbf{Dominant driven mode.} The mode resonantly excited by the stripline carries the largest coherent population. Secondary modes can
    therefore be treated as fluctuations about this driven background.
    \item \textbf{Weak nonlinear frequency scale.} The density-dependent
    nonlinear shifts and scattering rates, of order $|g|n$, remain small
    compared with the carrier frequencies of the participating magnons. This separation permits a perturbative expansion in the interaction.
    \item \textbf{Stationary continuous-wave response.} The microwave tones are applied continuously, and the measurement is performed after propagation and damping transients have decayed. The relevant coherent amplitudes may therefore be described by stationary spatial envelope equations.
\end{enumerate}

These approximations retain the energy- and momentum-matching conditions,
spatial propagation, and damping that determine the observed resonances, while separating the individual nonlinear scattering channels.

\subsection{Bogoliubov Decomposition}

Following the first simplifying aspect, we decompose the magnon field into the directly driven mode and fluctuations:
\begin{equation}
    \Psi_{\bm k}(t) = \psi_{\bm k}(t) + \delta \psi_{\bm k}(t),
\end{equation}
where $\psi_{\bm k}$ represents the directly driven mode and $\delta \psi_{\bm k}$ captures new modes excited through nonlinear interactions. Substituting this decomposition into \equ{eq:NLWE_momentum} and keeping terms to leading order in $\delta\psi$, we obtain for the driven mode:
\begin{equation}
    i \partial_t \psi_{\bm k}(t)  =  (  \omega_{\bm k}  -  i \Gamma_{\bm k} )  \psi_{\bm k} + \sum_{\bm k_1, \bm k_2 ,\bm k_3} g_{3k}^{12} \delta_{3k}^{12} \psi_1 \psi_2 \psi_3^*   + S_{\bm k}(t),
    \label{eq:NLWE_driven_mode}
\end{equation}
and for the fluctuations:
\begin{equation}
    i \partial_t \delta \psi_{\bm k}(t)  =  (  \omega_{\bm k}  -  i \Gamma_{\bm k} )  \delta \psi_{\bm k} + \sum_{\bm k_1, \bm k_2 ,\bm k_3} g_{3k}^{12} \delta_{3k}^{12} \left[  2 \delta \psi_1 \psi_2 \psi_3^*  + \psi_1 \psi_2 \delta \psi_3^* \right].
    \label{eq:NLWE_fluctuations}
\end{equation}

\subsection{Perturbative Expansion}

Even \equ{eq:NLWE_driven_mode} remains nonlinear. Following the second simplifying aspect, we treat the nonlinearity perturbatively. Introducing a formal expansion parameter $\epsilon$ ($g^{12}_{3k} \to \epsilon g^{12}_{3k}$), we expand:
\begin{equation}
    \psi_{\bm k} = \psi_{\bm k}^{(0)} + \epsilon \psi_{\bm k}^{(1)} + \epsilon^2 \psi_{\bm k}^{(2)} + \ldots
    \label{eq:perturbative_expansion}
\end{equation}
To zeroth order, interactions are absent:
\begin{equation}
    i\partial_t  \psi_{\bm k}^{(0)}(t) =  \left[ \omega_{\bm k} - i \Gamma_{\bm k} \right] \psi_{\bm k}^{(0)}(t)  + S_{\bm k}(t).
    \label{eq:NLWE_linear}
\end{equation}
For a general multi-tone drive of the form $S_{\bm k}(t) = \sum_j S_{\bm k_j} e^{-i\omega_j t}$, the steady-state solution is:
\begin{equation}
   \psi_{\bm k}^{(0)}(t) = \sum_j \frac{S_{\bm k_j}}{\omega_{\bm k} - \omega_j - i\Gamma_{\bm k}} e^{-i\omega_j t} \equiv \sum_j A_j^{(0)}(\bm k - \bm k_j) e^{-i\omega_j t},
   \label{eq:general_drive_solution}
\end{equation}
where $A_j^{(0)}(\bm k - \bm k_j)$ is the envelope function for the $j$-th drive tone, peaked around the resonant wavevector $\bm k_j$ satisfying $\omega(\bm k_j) = \omega_j$.

At first order, the driven modes interact to generate new frequency components:
\begin{equation}
    \left( i \partial_t - \omega_{\bm k} + i \Gamma_{\bm k}\right) \psi_{\bm k}^{(1)}(t) = \sum_{\bm k_1, \bm k_2, \bm k_3} g^{12}_{3k} \delta^{12}_{3k} \psi^{(0)}_1 \psi^{(0)}_2 \psi^{(0),*}_3.
    \label{eq:first_order_correction}
\end{equation}
The right-hand side contains products of three zeroth-order fields, generating frequencies that are sums and differences of the input frequencies.

\subsection{Application to Two-Tone Drive}

We now specialize to the two-tone experiment, where we apply simultaneous microwave drives at a pump frequency $\omega_p$ and a signal frequency $\omega_s$. The zeroth-order solution \equ{eq:general_drive_solution} becomes:
\begin{equation}
   \psi_{\bm k}^{(0)}(t) = A_p(\bm k-\bm k_p) e^{-i \omega_p t} + A_s(\bm k-\bm k_s) e^{-i \omega_s t},
   \label{eq:two_tone_solution}
\end{equation}
where the envelope functions $A_p$ and $A_s$ are determined by the injection profile derived in \secu{sec:magnon-interaction}, and the wavevectors satisfy the dispersion relation: $\omega_p = \omega(\bm k_p)$ and $\omega_s = \omega(\bm k_s)$.

\subsection{Derivation of the Idler Envelope Equation}

Substituting the two-tone solution \equ{eq:two_tone_solution} into the first-order equation \equ{eq:first_order_correction}, the right-hand side contains terms oscillating at frequencies $\omega_p$, $\omega_s$, $2\omega_p - \omega_s$, and $2\omega_s - \omega_p$. At this level, the labels ``pump'' and ``signal'' are completely interchangeable: both $2\omega_p - \omega_s$ and $2\omega_s - \omega_p$ are generated on equal footing. We focus on the idler frequency $\omega_i = 2\omega_p - \omega_s$, which in our experiment is tuned to match the NV resonance.

To derive the spatial evolution of the idler, we assume the envelope functions are slowly varying compared to the carrier wavelength. Writing the idler as $\psi_{\bm k}^{(1)}(t) = A_i(\bm k - \bm k_i) e^{-i\omega_i t}$, where $\bm k_i$ satisfies $\omega(\bm k_i) = \omega_i$, we expand the dispersion relation to linear order:
\begin{equation}
    \omega(\bm k) \approx \omega(\bm k_i) + \bm v_{\bm k_i} \cdot (\bm k - \bm k_i), \qquad \text{where} \qquad \bm v_{\bm k_i} = \nabla_{\bm k} \omega \big|_{\bm k = \bm k_i}
\end{equation}
is the group velocity at $\bm k_i$. Transforming to real space via the Fourier convention
\begin{equation}
     A(\bm k) = \frac{1}{L^2} \int \d \bm r \, A(\bm r)\,  e^{i \bm k \cdot \bm r}, \quad  A(\bm r) = \sum_{\bm k} A(\bm k) e^{-i \bm k \cdot \bm r},
\end{equation}
we obtain the envelope equation in steady state:
\begin{equation}
    \left[ \bm v_{\bm k_i} \cdot \nabla +  \Gamma_i\right] A_i(\bm r) = -i g^{pp}_{si} A_p^2(\bm r) A^*_s(\bm r) e^{i \bm{\Delta k}\cdot \bm r},
    \label{eq:idler_envelope_full}
\end{equation}
where $\Gamma_i = \Gamma(\bm k_i)$ is the damping rate at the idler wavevector, and $\bm{\Delta k} = 2 \bm k_p - \bm k_s - \bm k_i$ is the phase mismatch. This equation describes stimulated four-magnon scattering: the idler amplitude grows proportionally to $A_p^2 A_s^*$, representing the annihilation of two pump magnons to create a signal and an idler magnon, stimulated by the seeded signal mode.

\subsection{Solution for Damon--Eshbach Propagation}

In our geometry, magnons propagate predominantly along the $y$-direction (Damon--Eshbach configuration). Restricting to this direction, \equ{eq:idler_envelope_full} becomes:
\begin{equation}
    \left[  v^y_i \partial_y + \Gamma_i\right] A_i(y) = -i g^{pp}_{si} A_p^2(y) A^*_s(y) e^{i \Delta k^y y},
    \label{eq:idler_envelope_1D}
\end{equation}
where $v^y_i = \text{sgn}(k_i^y) \, v_g(k_i^y)$ is the exact $y$-component of the group velocity evaluated at the idler wavevector. 

Assuming the pump and signal envelopes decay exponentially due to damping:
\begin{equation}
    A_p(y) = A_p(0) e^{- \lambda_p y}, \qquad A_s(y) = A_s(0) e^{- \lambda_s y},
\end{equation}
where $\lambda_{p,s} = \Gamma_{p,s}/v^y_{p,s}$ are the spatial decay rates, \equ{eq:idler_envelope_1D} can be solved with the boundary condition $A_i(0) = 0$ (no idler at the stripline). Defining $\lambda_i = \Gamma_i/v^y_i$, $\lambda_0 = 2\lambda_p + \lambda_s$, and $A_0 = -ig^{pp}_{si} A_p^2(0) A_s^*(0)/v^y_i$, the solution is:
\begin{equation}
    A_i(y) = \frac{A_0}{\lambda_0 - \lambda_i - i\Delta k^y} \left[ e^{-\lambda_i y} - e^{(-\lambda_0 + i \Delta k^y)y}\right].
    \label{eq:idler_solution_full}
\end{equation}

For phase-matched propagation ($\Delta k^y = 0$) and negligible idler damping ($\lambda_i \ll \lambda_0$), this simplifies to:
\begin{equation}
    A_i(y) \approx \frac{A_0}{\lambda_0} \left[1 - e^{-\lambda_0 y} \right].
    \label{eq:idler_solution_simple}
\end{equation}
In the limit of weak pump/signal damping, the idler grows linearly with propagation distance: $A_i(y) \approx A_0 y$. This linear growth regime is characteristic of stimulated four-magnon scattering and underlies our detection scheme.

\subsection{Connection to Measured Signal}

The idler frequency $\omega_i = 2\omega_p - \omega_s$ is tuned to the NV resonance $\omega_{\rm NV}$, so the NV detects the idler alone; the pump and signal are far off-resonance and do not contribute. The idler envelope \equ{eq:idler_solution_full} therefore sets the local magnon population density $\langle \tilde n_i(y) \rangle = |A_i(y)|^2$, which maps to the measured Rabi frequency through \equ{eq:nq_from_Rabi_combined} of \secu{sec:magnon-interaction}, evaluated with the local separation $z_{\rm NV}$ from \secu{sec:height_mapping} and the idler wavevector $q_y$ fixed by the dispersion relation.

To extract the four-magnon interaction vertex $g^{pp}_{si}$, we fit the spatial profile of the measured idler population profile, $\langle \tilde n_i(y)\rangle = |A_i(y)|^2$, to the squared magnitude of \equ{eq:idler_solution_full}. The source amplitude $A_0 = -ig^{pp}_{si} A_p^2(0) A_s^*(0)/v^y_i$ contains the coupling constant $g^{pp}_{si}$ as an unknown parameter, while the pump and signal amplitudes $A_p(0)$, $A_s(0)$ are independently determined from single-tone Rabi measurements and the stripline current calibration (see \secu{sec:power}). {} This fitting procedure yields the extracted values of $g^{pp}_{si}$ shown in Fig.~2(F) of the main text.

\section{Cascaded Scattering Model}\label{sec:cascaded}

When driving the magnon system with a single tone at frequency $\omega_0$, the directly excited mode can become parametrically unstable above a threshold power, spontaneously generating pairs of magnons at new frequencies. These secondary modes can themselves become unstable, leading to a cascade of parametric resonances. Here we derive the conditions for this cascaded instability and show how it determines the observable resonance structure.

\subsection{Parametric Instability from Four-Magnon Scattering}

We begin with the Bogoliubov decomposition introduced in the previous section. For a single-tone drive, the zeroth-order solution is:
\begin{equation}
    \psi_{\bm k}^{(0)}(t) = A_0(\bm k - \bm k_0) e^{-i\omega_0 t},
\end{equation}
where $\omega_0$ is the drive frequency and $\bm k_0$ satisfies the dispersion relation $\omega(\bm k_0) = \omega_0$. Unlike the two-tone case where we treated the idler perturbatively, here we examine the stability of the driven mode against spontaneous decay into pairs of secondary modes.

The fluctuation equation \equ{eq:NLWE_fluctuations} describes how small perturbations evolve in the presence of the driven mode. We write the fluctuations as a superposition of two modes with frequencies $\omega_1$ and $\omega_2$ constrained by energy conservation ($\omega_1 + \omega_2 = 2\omega_0$):
\begin{equation}
    \delta \psi_{\bm k}(t) = A_1(\bm k - \bm k_1) e^{-i\omega_1 t} + A_2(\bm k - \bm k_2) e^{-i\omega_2 t}.
\end{equation}
Momentum conservation requires $\bm k_1 + \bm k_2 = 2\bm k_0$.

\subsection{Coupled Mode Equations}

Substituting into \equ{eq:NLWE_fluctuations} and using the slow-envelope approximation, we obtain coupled equations for the spatial evolution of the two modes. Including the mean-field frequency shift from the driven mode, the equations in steady state become:
\begin{align}
    \left[\omega_1 - \tilde{\omega}(\bm k_1) + i \bm v_{\bm k_1} \cdot \nabla + i\Gamma(\bm k_1)\right] A_1(\bm r) &= g^{00}_{12} A_0^2(\bm r) A_2^*(\bm r), \label{eq:coupled_mode_1} \\
    \left[\omega_2 - \tilde{\omega}(\bm k_2) - i \bm v_{\bm k_2} \cdot \nabla - i\Gamma(\bm k_2)\right] A_2^*(\bm r) &= g^{00}_{12} [A_0^*(\bm r)]^2 A_1(\bm r), \label{eq:coupled_mode_2}
\end{align}
where the renormalized dispersion relation includes the mean-field shift:
\begin{equation}
    \tilde{\omega}(\bm k) \approx \omega(\bm k) + 2g^{0k}_{0k} |A_0|^2.
    \label{eq:mean_field_shift}
\end{equation}
The structure of Eqs.~(\ref{eq:coupled_mode_1})--(\ref{eq:coupled_mode_2}) is characteristic of parametric amplification: the gain of mode $A_1$ is proportional to $A_2^*$, and vice versa. This mutual coupling enables exponential growth when the parametric drive exceeds a threshold.

\subsection{Spatial Growth Rate}

For modes propagating along the $y$-direction, we seek solutions of the form:
\begin{equation}
    A_1(y) = a_1 e^{\lambda y}, \qquad A_2(y) = a_2 e^{\lambda y},
\end{equation}
where $\lambda$ is the spatial growth rate. Substituting into the coupled mode equations yields the matrix equation:
\begin{equation}
    \begin{bmatrix}
       \omega_1 - \tilde{\omega}(\bm k_1) + iv^y_{\bm k_1}\lambda + i\Gamma(\bm k_1) & -g^{00}_{12} [A_0^*]^2 \\[6pt]
       -g^{00}_{12} A_0^2 & \omega_2 - \tilde{\omega}(\bm k_2) - iv^y_{\bm k_2}\lambda - i\Gamma(\bm k_2)
    \end{bmatrix}
    \begin{bmatrix}
        a_1 \\
        a_2^*
    \end{bmatrix}
    = 0.
    \label{eq:parametric_matrix}
\end{equation}

For non-trivial solutions, the determinant must vanish. Defining the frequency-momentum mismatch parameters:
\begin{align}
    v^y_{\bm k_1} q_1 &= \omega_1 - \tilde{\omega}(\bm k_1), \\
    v^y_{\bm k_2} q_2 &= \omega_2 - \tilde{\omega}(\bm k_2),
\end{align}
the solvability condition in the absence of damping gives:
\begin{equation}
    \lambda = -\frac{i}{2}(q_1 - q_2) \pm \frac{1}{2}\sqrt{\frac{4|g^{00}_{12} A_0^2|^2}{v^y_{\bm k_1} v^y_{\bm k_2}} - (q_1 + q_2)^2}.
    \label{eq:growth_rate_general}
\end{equation}
For the growth rate to be real (indicating true instability rather than oscillation), we require the expression under the square root to be positive. Setting $q_1 = q_2 = q$ for maximum growth, where:
\begin{equation}
    q = \frac{\tilde{\omega}(\bm k_1) + \tilde{\omega}(\bm k_2) - 2\omega_0}{v^y_{\bm k_1} + v^y_{\bm k_2}},
\end{equation}
the growth rate simplifies to:
\begin{equation}
    \lambda = \pm\sqrt{\frac{|g^{00}_{12} A_0^2|^2}{v^y_{\bm k_1} v^y_{\bm k_2}} - q^2}.
    \label{eq:growth_rate_simplified}
\end{equation}

\subsection{Selection Rules for Unstable Modes}

Equation~(\ref{eq:growth_rate_simplified}) reveals important constraints on which modes can become unstable:

\begin{enumerate}
    \item \textbf{Group velocity sign constraint:} The product $v^y_{\bm k_1} v^y_{\bm k_2}$ must be positive for $\lambda$ to be real. Since the Damon--Eshbach group velocity $v^y_{\bm k} = \text{sgn}(k^y) v_{\rm DE}$, this requires both $k_1^y$ and $k_2^y$ to have the same sign. Combined with momentum conservation $k_1^y + k_2^y = 2k_0^y$, this restricts the allowed range to $0 \leq k_1^y \leq 2k_0^y$ (assuming $k_0^y > 0$).
    \item \textbf{Boundary dominance:} The growth rate diverges as $v^y_{\bm k} \to 0$, which occurs at $k^y = 0$ or $k^y = 2k_0^y$. Consequently, the most unstable modes lie along these boundaries in momentum space, corresponding to one of the daughter magnons having near zero $y$-momentum.
    \item \textbf{Energy conservation competition:} The detuning term $q^2$ penalizes modes that are off-shell. The most unstable mode balances the large growth from small group velocities against the cost of energy mismatch.
\end{enumerate}

\subsection{Cascade Hierarchy}

Once the primary instability generates modes at $\bm k_1$ and $\bm k_2$, these can themselves act as ``pump'' modes for subsequent parametric processes. The cascade proceeds hierarchically: each scattering step feeds the next.

Since the most unstable modes have one daughter at $k^y = 0$ or $k^y = 2k_{\rm in}^y$, the cascade preferentially populates modes along specific lines in momentum space. For a drive at $\bm k_0$, the dominant cascade path places one daughter at $k^y = 0$ while the other carries all the $y$-momentum. Successive steps thus produce modes at $k^y = 2k_0^y, 3k_0^y, \dots$, and we label each resonance by its cascade order $n$ (the same index used in the main text), defined such that the order-$n$ process involves the mode at $k^y = n\,k_0^y$. The primary instability corresponds to $n = 2$.

The NV center detects magnons resonant at $\omega_{\rm NV}$. The order-$n$ cascade reaches the NV resonance when:
\begin{equation}
    \omega\big(n\,\bm k_0\big) = \omega_{\rm NV}.
\end{equation}

\subsection{Mean-Field Frequency Shift and Resonance Condition}

The mean-field shift \equ{eq:mean_field_shift} modifies the resonance condition. For the linearized Damon--Eshbach dispersion $\omega_{\bm k} = \Delta_0 + v_{\rm DE}|k^y|$, the renormalized dispersion becomes:
\begin{equation}
    \tilde{\omega}_{\bm k} = \Delta_0 + 2\omega_{\rm NL} + v_{\rm DE}|k^y|,
\end{equation}
where $\omega_{\rm NL} \approx g^{0k}_{0k} n_0 \propto |A_0|^2$ depends on the pump magnon density. Because the driven magnon amplitude scales linearly with the stripline current ($A_0 \propto I$; see \secu{sec:power}), the pump density is quadratic in the current, $n_0 = |A_0|^2 \propto I^2$, and the mean-field shift is therefore quadratic, $\omega_{\rm NL} \propto I^2$. This justifies the parameterization $\omega_{\rm NL} = \beta I^2$ used below.

For the order-$n$ cascade to reach the NV resonance, the drive frequency must satisfy:
\begin{equation}
    \omega_0(n, \omega_{\rm NL}) = \frac{\omega_{\rm NV} + (n-1)\Delta_0}{n} + \frac{2(n-1)}{n}\omega_{\rm NL}.
    \label{eq:resonance_condition}
\end{equation}
This equation predicts that the resonance frequencies shift with drive power (through $\omega_{\rm NL}$), providing a distinctive experimental signature of the cascaded scattering mechanism.

\subsection{Fitting the Cascade Resonances}

\begin{figure*}[t]
    \centering
    \includegraphics[width=\linewidth]{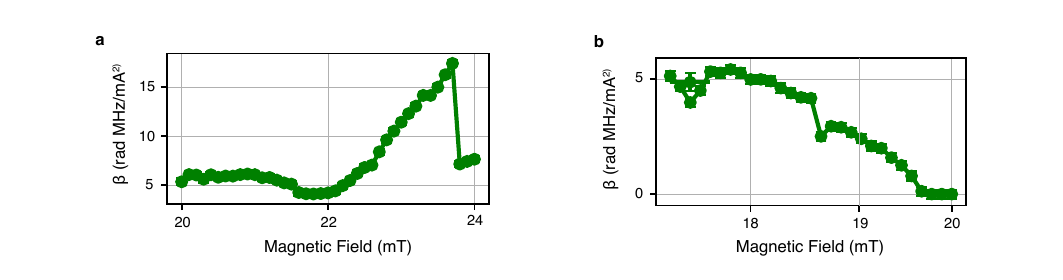}
    \caption{
        \textbf{Mean-field frequency shifts of the cascade resonances.}
        Fitted parameters of the resonance condition \equ{eq:resonance_condition}, extracted from the power-dependent frequency shift of the cascade resonances. For each applied magnetic field $H_{\rm ext}$, we plot the mean-field shift coefficient $\beta$ (defined via $\omega_{\rm NL} = \beta I^2$) together with the fitted cascade order $n$.
        (a) Results for the resonance shown in Fig.~3(C)(i) of the main text. The fitted cascade order $n = 1.51(\pm 0.91)$ is consistent with $n = 2$.
        (b) Results for the resonance shown in Fig.~3(C)(ii) of the main text. The fitted cascade order $n = 2.85(\pm 0.77)$ is consistent with $n = 3$.
    }
    \label{fig:figureS3}
\end{figure*}

To fit the observed resonances, we use \equ{eq:resonance_condition} with $\omega_{\rm NL} = \beta I^2$, where $I$ is the stripline current and $\beta$ is a fitting parameter that encodes the magnon-magnon interaction strength and the relation between current and magnon density. The cascade order $n$ determines the slope of the power-dependent frequency shift, allowing us to identify which cascade order produces each observed resonance. The predicted resonance structure, discrete peaks that shift with power according to their cascade order, matches the experimental observations and confirms the cascaded parametric scattering mechanism. The fitted mean-field coefficient $\beta$ and cascade order $n$ for each resonance are shown in \figu{fig:figureS3}. 

\subsection{Numerical Prediction of the Resonance Map}\label{sec:arnold_tongues}

The analytical results derived above including the parametric growth rate (\equ{eq:growth_rate_simplified}), the cascade hierarchy, and the resonance condition (\equ{eq:resonance_condition}) predict that specific drive frequencies $\omega_0$ will produce magnons at the NV resonance $\omega_{\rm NV}$ for each applied field $H_{\rm ext}$. To obtain a quantitative prediction of the full resonance structure shown in Fig.~3 of the main text, we numerically evaluate the cascaded growth rate and its overlap with the NV momentum filter across a two-dimensional grid of $(H_{\rm ext}, \omega_0)$ values. The bright ridges in the resulting color map trace out an resonance structure, identifying the drive frequencies at which parametric scattering most efficiently channels magnon population into the NV detection window.

\subsubsection{Drive momentum and field-dependent parameters}

The calculation sweeps over magnetic fields $\mu_0 H_{\rm ext} \in [17, 24]\,\text{mT}$ and drive frequencies $\omega_0/(2\pi) \in [1.75, 2.10]\,\text{GHz}$. For each $H_{\rm ext}$, we compute the equilibrium magnetization angle $\theta_0$ and the full dispersion relation. The NV resonance frequency is
\begin{equation}
    \omega_{\rm NV}(H_{\rm ext}) = 2\pi \times 2.87\,\text{GHz} - \gamma H_{\rm ext}\,.
\end{equation}
For each drive frequency $\omega_0$, the corresponding drive momentum $\bm k_0 = (0, k_0^y)$ is found by numerically inverting the one-dimensional Damon--Eshbach dispersion relation along $k_y$: we find $k_0^y$ such that $\omega(0, k_0^y) = \omega_0$. The constraint $k_0^x = 0$ follows from the stripline injection profile derived in \secu{sec:magnon-interaction}, which confines excitation to the $k_x = 0$ subspace. When the mean-field shift is included, the inversion uses the renormalized dispersion $\tilde\omega(\bm k) = \omega(\bm k) + 2\omega_{\rm NL}$.

\subsubsection{Growth rate on the two-dimensional momentum grid}

At each $(H_{\rm ext}, \omega_0)$ point, the parametric growth rate from \equ{eq:growth_rate_simplified} is evaluated over a two-dimensional momentum grid spanning $k_x \in [-40, 40]\,\upmu\text{m}^{-1}$ and $k_y \in [-0.1, 4]\,\upmu\text{m}^{-1}$ with $10^3 \times 10^3$ points. Each grid point represents a candidate daughter momentum $\bm k_3$; the partner is fixed by momentum conservation, $\bm k_4 = \bm k_1 + \bm k_2 - \bm k_3$, where $\bm k_1$ and $\bm k_2$ are the two input momenta for the scattering process. The growth rate is
\begin{equation}
    \lambda(\bm k_3) = \begin{cases}
    \displaystyle\sqrt{\frac{|g^{12}_{34}\, A_{1}\, A_{2}|^2}{v^y_{\bm k_3}\, v^y_{\bm k_4}} - q^2} & \text{if the argument is positive}\,,\\[8pt]
    0 & \text{otherwise}\,,
    \end{cases}
    \label{eq:lambda_numerical}
\end{equation}
where the energy--momentum mismatch is
\begin{equation}
    q = \frac{\tilde\omega(\bm k_3) + \tilde\omega(\bm k_4) - \omega_{\bm k_1} - \omega_{\bm k_2}}{v^y_{\bm k_3} + v^y_{\bm k_4}}\,,
\end{equation}
and $A_{1}$, $A_{2}$ and $\omega_{\bm k_1}$, $\omega_{\bm k_2}$ are the amplitudes and frequencies of the two incoming modes. The group velocities are evaluated using the exact numerical derivative $v^y_{\bm k} = \operatorname{sgn}(k^y)\, \partial \omega / \partial k$ to properly account for the flattening of the dispersion curve. The bare four-magnon interaction strength is taken to be momentum-independent and normalized by the effective macroscopic spin density $s_{\rm eff}$:
\begin{equation}
    g \approx -\frac{\gamma M_s}{2 s_{\rm eff}}\,,
    \label{eq:g_momentum_independent}
\end{equation}
corresponding to the dominant dipolar contribution to the T-matrix and evaluates to $g/(2 \pi) \approx -0.3 \; \text{Hz}\, \upmu \text{m}^{3}$. This simplification replaces the full momentum-dependent vertex $g^{12}_{34}$ in \equ{eq:lambda_numerical} with the single value $g$. Using the full momentum-dependent vertex shifts the position of the maximum only marginally, since the group-velocity enhancement and off-shell penalty provide the dominant momentum dependence.

\subsubsection{Cascade through four scattering steps}

The cascade is computed iteratively: the most unstable output momenta from each step serve as inputs for the next. The branching structure proceeds as follows:
\begin{enumerate}
    \item \textbf{Step 1} (order $n=2$): Two pump magnons scatter, $\bm k_0 + \bm k_0 \to \bm k_1 + \bm k_2$. This produces one growth rate map and identifies the most unstable pair $(\bm k_1^{\rm max}, \bm k_2^{\rm max})$.
    
    \item \textbf{Step 2} (order $n=3$): The pump scatters with each first-step daughter, $\bm k_0 + \bm k_1^{\rm max} \to \bm k_3 + \bm k_4$ and $\bm k_0 + \bm k_2^{\rm max} \to \bm k_3' + \bm k_4'$, producing 2 growth rate maps.
    
    \item \textbf{Step 3} (order $n=4$): Four standard branches arise from the pump scattering with each second-step output. Two additional branches exploit the dispersion relation symmetry under $k_x \to -k_x$: the first-step daughter $\bm k_1^{\rm max}$ scatters with its mirror image $\bm k_{1,\rm mirror} = (-k_{1x}^{\rm max},\, k_{1y}^{\rm max})$, and similarly for $\bm k_2^{\rm max}$. This yields 6 branches total.
    
    \item \textbf{Step 4} (order $n=5$): Each of the 12 output momenta from step 3 scatters with the pump, producing 12 branches.
\end{enumerate}
At each generation, we retain only the momentum that maximizes the growth rate on each parent branch. Because the higher-order mode populations are not independently known, all input modes are assigned the same momentum-independent reference amplitude, $A_{\rm ref}=\sqrt{n_0}$. This choice does not shift the momentum maximizing $\lambda(\bm k)$ over the parameter range considered. The calculation therefore predicts the resonance positions and branching structure, not their spectral weights.

\subsubsection{Overlap with the NV detection window}

For each cascade order and branch, the predicted NV signal strength is computed as the overlap between the growth rate map and the NV-detectable region of momentum space. First, the NV-resonant momenta are identified as the set of wavevectors $\bm k$ on the 2D grid satisfying $|\omega(\bm k) - \omega_{\rm NV}| < \delta\omega$, forming an iso-energy contour in the $(k_x, k_y)$ plane. Each point on this contour is then weighted by the NV momentum filter function $\tilde{\mathcal{D}}^{\downarrow,+-}_{\bm q}$ from \equ{eq:mom_filter_T1}, which encodes the dipolar coupling strength between magnons at wavevector $\bm q$ and the NV center. The filter captures the essential physics: the exponential suppression $e^{-2q z_{\rm NV}}$ of coupling to large wavevectors, and the angular selection rules set by the NV and magnetization orientations. {}

The overlap for a given branch at order $n$ is
\begin{equation}
    \mathcal{O}_n(H_{\rm ext}, \omega_0) = \sum_{\bm k \,\in\, \text{NV contour}} \lambda^{(n)}(\bm k)\; \hat{\mathcal{D}}(\bm k)\,,
    \label{eq:NV_overlap}
\end{equation}
where $\hat{\mathcal{D}}(\bm k)$ denotes the normalized and thresholded momentum filter. This scalar value quantifies the efficiency with which the order-$n$ cascade delivers parametrically amplified magnons into the NV detection window at the given $(H_{\rm ext}, \omega_0)$ operating point.

\subsubsection{Assembly and comparison with experiment}

The overlap values $\mathcal{O}_n(H_{\rm ext}, \omega_0)$ are computed across the full parameter space, and the contributions from all branches within each order are summed. The resulting four two-dimensional maps, one per cascade order, constitute the resonance diagram. The bright ridges, where $\mathcal{O}_n$ is maximal, predict the drive frequencies at which the order-$n$ cascade produces magnons at $\omega_{\rm NV}$.

Two cases are considered for comparison with experiment: for the lower cascade orders ($n = 2, 3$), the mean-field shift is set to zero ($\omega_{\rm NL} = 0$), appropriate for moderate pump densities where the power-dependent dispersion correction is small. For the higher orders ($n = 4, 5$), where the driven magnon density is large enough to appreciably shift the dispersion relation, a finite mean-field correction $\omega_{\rm NL} = g\, n_0$ is included. The predicted resonance map is compared with the experimental observations in Fig.~3 of the main text, where the predicted tongue maxima align with the experimentally measured resonance frequencies across the full range of magnetic fields.

\section{Power Calibration}\label{sec:power}

\subsection{Stripline Current Calibration}

Throughout this work, the microwave stripline current $I$ enters the theory as the source of the magnetic field that drives magnons in the YIG film. In particular, the injection amplitude $S_0 \propto I$ derived in \secu{sec:magnon-interaction} sets the driven magnon population, and the pump amplitude $|A_0|^2 \propto I^2$ determines the parametric gain in \secu{sec:cascaded}. Here we describe how $I$ is determined from an experimentally accessible quantity: the microwave power measured at the output of the circuit.

\subsubsection{Microwave circuit and measurement}

As described in the main text, the microwave circuit consists of a signal generator (50~$\Omega$ output impedance), connected via coaxial cables and a PCB trace to the gold stripline antenna on the YIG chip. One end of the stripline is wire bonded to the PCB; the other end is wire bonded back to a second PCB trace, which is connected to a 50~$\Omega$ termination, which is a Signal Hound spectrum analyzer. Both terminations present a 50~$\Omega$ load. The spectrum analyzer records the transmitted power $P_{\rm SA}$ in dBm, from which the current at the output port is obtained as
\begin{align}
    I_{\rm SA} = \sqrt{\frac{2 P_{\rm SA}}{Z_0}}\,, \qquad Z_0 = 50\;\Omega\,.
    \label{eq:ISA_from_PSA}
\end{align}
This is the current flowing through the input impedance of the spectrum analyzer. We now argue that $I_{\rm SA}$ is a faithful measure of the stripline current $I$.

\subsubsection{Lumped-element argument}

The gold stripline has a length of approximately 400~$\mu$m. At the operating frequencies of 1.5--3~GHz, the electromagnetic wavelength is on the order of centimeters, exceeding the stripline length by two orders of magnitude. The stripline is therefore electrically short and acts as a lumped series element (predominantly a small inductance). The same applies to the wire bonds, which contribute a small series inductance.

In this lumped-element limit, the entire path from one PCB pad through the wire bond, stripline, and second wire bond to the opposite PCB pad is a series network. In a series circuit, the current is the same at every point. The current flowing through the stripline is therefore equal to the current flowing into the 50~$\Omega$ termination:
\begin{align}
    I_{\rm strip} \approx I_{\rm SA}\,.
    \label{eq:Istrip_approx_ISA}
\end{align}

To verify this picture, we measure the forward transmission coefficient $|S_{21}|^2$ of the full circuit (from signal generator output to spectrum analyzer input) as a function of frequency across the experimental bandwidth. The transmission is found to be approximately flat, with no significant resonant features.

\subsection{Calculating the Pump and Signal Magnon Amplitudes}\label{sec:pump_signal_calibration}

The fitting procedure described in \secu{sec:stimulated} requires knowledge of the pump and signal magnon population densities $\langle \tilde{n}_p \rangle$ and $\langle \tilde{n}_s \rangle$ at the stripline. These cannot be measured directly from the NV Rabi signal, as the pump and signal are far off-resonance with the NV transition, but can be inferred from single-tone, Rabi measurements on resonance with the NV transition ($\omega_{\rm NV} = \omega_{\rm drive}$) combined with the stripline current calibration described above.

\subsubsection{Reference calibration from single-tone Rabi measurements}

In a single-tone experiment, the stripline drives magnons at a single frequency $\omega_d$ with current $I$. At the applied bias field, the NV resonance condition $\omega_{\rm NV} = \Delta -  \gamma H^z$ selects a magnon wavevector $k_{\rm NV}$ via the dispersion relation. These magnons propagate away from the stripline, and the NV center detects the resulting stray field as a Rabi frequency $\Omega_R$.

Using the macroscopic mapping equations derived in \secu{sec:magnon-interaction} and the NV--YIG separation $z_{\rm NV}$ from the height map (\secu{sec:height_mapping}), we convert the measured Rabi frequency directly to a local magnon population density $\langle \tilde{n}_{\rm NV} \rangle$. This provides a reference calibration point: at current $I_{\rm ref}$ and the reference wavevector $k_{\rm ref} \equiv k_{\rm NV}$ of this measurement, we obtain a measured population density $\langle \tilde{n}_{\rm ref} \rangle$.

\subsubsection{Pump and signal population densities}

We calculate the current$\,\to\,$magnon transduction directly as follows. Sweeping the bias field tunes the NV-resonant wavevector $k_{\rm NV}$ through the dispersion relation, and the per-current Rabi response defines the normalized transduction
\begin{align}
    R(k) \equiv \frac{\Omega_R(k)/I}{\Omega_R(k_{\rm ref})/I_{\rm ref}} \, ,
    \label{eq:filter_function}
\end{align}
i.e., the relative current$\,\to\,$magnon conversion efficiency at wavevector $k$, normalized to the reference measurement $\langle \tilde{n}_{\rm ref} \rangle$ above. We fit the experimentally-determined $R(k)$ to an exponentially decaying envelope, which we evaluate at the pump and signal wavevectors below. We can approximate $R(k) \approx \tilde{S}(k)/\tilde{S}(k_{\rm ref})$, with $\tilde{S}(k) = h(kH)w(kW)$ being the injection profile of \secu{sec:magnon-interaction}.

For the two-tone experiment, the pump (frequency $\omega_p$, current $I_p$) and signal (frequency $\omega_s$, current $I_s$) excite modes at wavevectors $k_p$ and $k_s$. Since the magnon amplitude scales linearly with current and the density is quadratic in the amplitude, $\langle \tilde n\rangle = |A|^2$, the local densities at the NV location are
\begin{align}
    \langle \tilde{n}_p(y_{\rm NV}) \rangle &= \langle \tilde{n}_{\rm ref} \rangle \left(\frac{I_p}{I_{\rm ref}}\right)^2 R(k_p)^2 \, , \label{eq:n_pump} \\
    \langle \tilde{n}_s(y_{\rm NV}) \rangle &= \langle \tilde{n}_{\rm ref} \rangle \left(\frac{I_s}{I_{\rm ref}}\right)^2 R(k_s)^2 \, . \label{eq:n_signal}
\end{align}
Because the scattering equations of \secu{sec:stimulated} are integrated forward from the stripline at $y=0$, these boundary amplitudes are back-propagated from the NV location $y_{\rm NV}$ to the source through their spatial decay lengths $\lambda = \Gamma/v_g$:
\begin{align}
    |A_p(0)|^2 &= \langle \tilde{n}_p(y_{\rm NV}) \rangle \, e^{2 \lambda_p y_{\rm NV}} \, , \\
    |A_s(0)|^2 &= \langle \tilde{n}_s(y_{\rm NV}) \rangle \, e^{2 \lambda_s y_{\rm NV}} \, .
\end{align}

\bibliography{paper_citations}